\documentclass[aps,prb,twocolumn,showpacs]{revtex4-1}

\usepackage{amsmath, amsthm, amssymb}
\usepackage{graphicx,color,psfrag}
\usepackage[normalem]{ulem}
\usepackage{subcaption}

\usepackage{color}

\newcommand{\Tr}[1]{\operatorname{Tr} \left\{ #1 \right\}}

\newcommand{\HOLU}{{\sc HOMO-LUMO} }

\DeclareMathAlphabet{\gcal}{OMS}{cmsy}{m}{n}
\newcommand{\myT}{{\gcal T}}

\graphicspath{{./AutipFigures/}}

\begin{document}

\title{Cold spots in quantum systems far from equilibrium: local entropies and temperatures near absolute zero}
\author{Abhay Shastry}
\affiliation{Department of Physics, University of Arizona, 1118 East Fourth Street, Tucson, AZ 85721}
\author{Charles\ A.\ Stafford}
\affiliation{Department of Physics, University of Arizona, 1118 East Fourth Street, Tucson, AZ 85721}
\date{\today}
\begin{abstract}
We consider a question motivated 
by the third law of thermodynamics: can there be a local temperature 
arbitrarily close to absolute zero in a nonequilibrium quantum system? 
We consider nanoscale quantum conductors with the source reservoir held at 
finite temperature and the drain held at or near absolute zero, a problem 
outside the scope of linear response theory. We obtain local temperatures close 
to absolute zero when electrons originating from the finite temperature 
reservoir undergo destructive quantum interference. 
The local temperature is computed by numerically solving a nonlinear system of 
equations describing equilibration of a scanning thermoelectric probe with the 
system, and we
obtain excellent agreement with analytic results derived using the Sommerfeld expansion. 
A local entropy
for a nonequilibrium quantum system is introduced, and used as a metric quantifying the departure from local equilibrium.
It is shown that the local entropy of the system tends to zero when the probe temperature tends to zero, consistent with the third law of
thermodynamics.
\end{abstract}


\maketitle

\section{Introduction}
\label{sec:intro}

The local temperature of a quantum system out of equilibrium is a concept of fundamental interest in nonequilibrium thermodynamics.
Out of equilibrium, the temperatures of different degrees of freedom generally do not coincide, so that one must distinguish between 
measures of lattice temperature
\cite{Chen03,Ming10,Galperin07}, 
photon temperature \cite{deWilde06,Yue11,Greffet07}, 
and electron temperature
\cite{Engquist81,Dubi09c,Sanchez11,Jacquet11,Caso11,Galperin11,Bergfield13demon,Meair14,Bergfield15,Stafford2014}.
The Scanning Thermal Microscope (SThM) \cite{Majumdar99} couples to all these degrees of freedom, and thus measures some linear combination of
their temperatures \cite{Bergfield15} in the linear response regime.
Recent advances in thermal microscopy
\cite{Kim11,Yu11,Kim12,Fabian12} have dramatically increased the spatial and thermal resolution of SThM, pushing it close to the quantum regime.

In this article, we focus on the local electron temperature $T_p$ as defined by a floating thermoelectric probe 
\cite{Bergfield13demon,Meair14,Bergfield15,Stafford2014}.  The probe, consisting of a macroscopic reservoir of electrons,
is coupled locally and weakly via tunneling to the system of interest,
and allowed to exchange charge and heat with the system until it reaches equilibrium, thus defining a simultaneous temperature and voltage measurement. 
Several variations on this measurement scenario have also been discussed in the literature \cite{Engquist81,Dubi09c,Sanchez11,Jacquet11,Caso11,Galperin11}.

The above definition of $T_p$ 
is operational: 
temperature is that which is measured by a suitably defined thermometer. Nonetheless,
it has been shown that this definition of $T_p$ 
is consistent with the laws of thermodynamics under certain specified conditions \cite{Meair14,Stafford2014}.
In the present article, we show that $T_p$ {\em is consistent with the third law of thermodynamics}.  In particular, we introduce a definition of
the local entropy $S_s$ of a nonequilibrium quantum system, and show that $S_s\rightarrow 0$ as $T_p\rightarrow 0$.  Moreover, we show that 
values of $T_p$ arbitrarily close to absolute zero can exist in quantum systems under thermal bias.  $S_s$ is also used 
to quantify the departure from local equilibrium beyond the linear response regime.

The article is organized as follows:  The nonequilibrium Green's function (NEGF) formalism needed to describe the local properties of
a nonequilibrium quantum system and its interaction with a scanning thermoelectric probe is introduced in Sec.\ \ref{sec:formalism}.
The local entropy of a nonequilibrium quantum system is defined in Sec.\ \ref{sec:entropy}, along with a normalization that takes into account
spatial variations in the local density of states.  Analytical results for the minimum local temperatures in quantum systems under thermal bias are
derived in Sec.\ \ref{sec:Tp_near_zero}.  The local temperature and entropy distributions in several $\pi$-conjugated molecular junctions under thermal
bias are computed in Sec.\ \ref{sec:results}.  Our conclusions are summarized in Sec.\ \ref{sec:conclusions}, while some useful details of the
formalism and modeling are provided in Appendices \ref{appendLocalEntropy}--\ref{append:ExactSolution}.

\section{Formalism}
\label{sec:formalism}

We consider a temperature/voltage probe
coupled locally (e.g., via tunneling) to a nonequilibrium quantum system.
The probe is also connected to an external macroscopic reservoir of noninteracting electrons held at a fixed temperature $T_p$ and chemical potential $\mu_p$.
We use the NEGF 
formalism to write the electron number current and heat current flowing into the probe as
\begin{equation}
\begin{aligned}
I_{p}^{(\nu)}&=\frac{-i}{h}\int_{-\infty}^{\infty}d\omega(\omega-\mu_{p})^{\nu}\\
&\Tr{\Gamma^{p}(\omega)\big(G^{<}(\omega)+f_{p}(\omega)[G^{r}(\omega)-G^{a}(\omega)])},
\label{GeneralFormula}
\end{aligned}
\end{equation}
where $\nu=0$ refers to the electron number current\, \cite{Meir92} and $\nu=1$ gives the electronic contribution to the heat current\, \cite{Bergfield09b}. 
$G^{r}(\omega)$ and $G^{a}(\omega)$
are the Fourier transforms of the retarded and advanced Green's functions, respectively, 
describing 
propagation of electronic excitations within the system, 
and $G^{<}(\omega)$ is the Fourier transform of the 
Keldysh ``lesser'' Green's function describing the nonequilibrium
population of the electronic spectrum of the system.
$\Gamma^{p}(\omega)$ is the tunneling-width
matrix describing the coupling of the probe to the system and $f_{p}(\omega)=1/(1+\exp{(\frac{\omega-\mu_{p}}{k_{B}T_{p}})})$ is the equilibrium Fermi-Dirac distribution of the probe.
Eq.\ (\ref{GeneralFormula}) is a general result valid for any interacting nanostructure under steady-state conditions.

\subsection{Local temperature and voltage measurements}
\label{sec:Tp_def}

A definition for a local electron temperature and voltage measurement on the system
that takes into account the thermoelectric corrections was proposed in Ref.\ \onlinecite{Bergfield13demon} by noting that the temperature $T_{p}$ and chemical 
potential $\mu_{p}$ should be simultaneously defined by the requirement that both the electric current and the electronic heat current into the probe vanish:
\begin{equation}
I_{p}^{(\nu)}=0,\ \ \ \ \nu \in \{0,1\}.
\label{equilibrium}
\end{equation}
Eq.\ (\ref{equilibrium}) gives the conditions under
which the probe is in local equilibrium with the sample, which is itself arbitrarily far from equilibrium.

Previous analyses \cite{Bergfield13demon,Meair14,Bergfield14,Bergfield15} have considered this problem within linear response theory, which reduces the system of nonlinear equations (\ref{equilibrium}) to
a system of equations linear in $T_{p}$ and $\mu_{p}$. In this article, we consider a problem that is essentially outside the linear response regime
and solve the nonlinear system of equations (\ref{equilibrium}) numerically.

It was shown in Ref.\ \onlinecite{Stafford2014} that Eq.\ (\ref{GeneralFormula}) can be written in terms of the local properties of the nonequilibrium system.
The mean local spectrum sampled by the probe was defined as
\begin{equation}
\bar{A}(\omega)\equiv\frac{\Tr{\Gamma^{p}(\omega)A(\omega)}}{\Tr{\Gamma^{p}(\omega)}},
\label{meanlocalspec}
\end{equation}
where $A(\omega)={i}\big(G^{r}(\omega)-G^{a}(\omega)\big)/2\pi$ is the spectral function of the nonequilibrium system.
Motivated by the relation at equilibrium, $G^{<}_{eq}(\omega)=2{\pi}iA(\omega)f_{eq}(\omega)$,
the local nonequilibrium distribution function (sampled by the probe) was defined as
\begin{equation}
f_{s}(\omega)\equiv\frac{\Tr{\Gamma^{p}(\omega)G^{<}(\omega)}}{2{\pi}i\Tr{\Gamma^{p}(\omega)A(\omega)}}.
\label{nonequilibriumdistribution}
\end{equation}
The mean local occupancy of the system orbitals sampled by the probe is \cite{Stafford2014}
\begin{equation}
\langle N\rangle 
=\int_{-\infty}^{\infty}d\omega\bar{A}(\omega)f_{s}(\omega),
\end{equation}
and similarly, the mean local energy of the system orbitals sampled by the probe is \cite{Stafford2014}
\begin{equation}
\langle E\rangle 
=\int_{-\infty}^{\infty}d\omega\omega\bar{A}(\omega)f_{s}(\omega).
\end{equation}
Eqs.\ (\ref{meanlocalspec}--\ref{nonequilibriumdistribution}) allow 
us to rewrite Eq.\ (\ref{GeneralFormula}) in a form analogous to the two-terminal Landauer-B\"{u}ttiker formula
\begin{equation}
\begin{aligned}
I_{p}^{(\nu)}&=\frac{1}{h}\int_{-\infty}^{\infty}d\omega(\omega-\mu_{p})^{\nu}\\
&2\pi\Tr{\Gamma^{p}(\omega)A(\omega)}[f_{s}(\omega)-f_{p}(\omega)].
\label{Rearraged}
\end{aligned}
\end{equation}
It was noted that, for the case of maximum local coupling, $[\Gamma^{p}(\omega)]_{ij}=\Gamma^{p}(\omega)\delta_{in}\delta_{jn}$,
the quantities $\bar{A}(\omega)=A_{nn}(\omega)$ and $f_{s}(\omega)$
become independent of the probe coupling, and can be related by the familiar equilibrium-type formula, $G^{<}_{nn}(\omega)=2{\pi}iA_{nn}(\omega)f_{s}(\omega)$,
even though the system is out of equilibrium.

In the broad-band limit $\Gamma^{p}(\omega)\approx\Gamma^{p}(\mu_{0})$, where $\mu_0$ is the equilibrium Fermi energy of the system,
we may write $\Tr{\Gamma^{p}(\mu_{0})}=\bar{\Gamma}^{p}$
so that $\bar{A}(\omega)=\Tr{\Gamma^{p}(\mu_{0})A(\omega)}/\bar{\Gamma}^{p}$. From Eq.\ (\ref{Rearraged}), we have
\begin{equation}
I_{p}^{(\nu)}=\frac{\bar{\Gamma}^{p}}{\hbar}\int_{-\infty}^{\infty}d\omega(\omega-\mu_{p})^{\nu}\bar{A}(\omega)[f_{s}(\omega)-f_{p}(\omega)].
\label{Broadband}
\end{equation}
It was noted that the equilibrium condition of Eq.\ (\ref{equilibrium}) now implies that the mean local occupancy
and energy of the nonequilibrium system 
are the same as if its nonequilibrium spectrum $\bar{A}(\omega)$ were populated by the equilibrium Fermi-Dirac distribution of 
the probe $f_{p}(\omega)$:
\begin{align}
\langle N\rangle\big|_{f_{p}}&=\langle N\rangle\big|_{f_{s}}
\label{BBOccupancy}\\
\langle E\rangle\big|_{f_{p}}&=\langle E\rangle\big|_{f_{s}},
\label{BBEnergy}
\end{align}
i.e., the probe equilibrates with the system in such a way that $f_{p}(\omega)$ satisfies the constraints imposed by
Eqs.\ (\ref{BBOccupancy}) and (\ref{BBEnergy}).

\subsection{Local entropy}
\label{sec:entropy}

The von Neumann entropy \cite{vonNeumann} for a system of noninteracting fermions can be expressed
in terms of the single-particle occupation probabilities $p_{i}$ as
\begin{equation}
S= -\sum_{i}[p_{i}\ln{p_{i}}+(1-p_{i})\ln{(1-p_{i})}],
\end{equation}
which can be extended to the case
of a continuous spectrum by simply replacing the summation by an integral over the density of states.
We propose a natural extension for the ``local entropy" $S_{s}$ of the nonequilibrium system, within an effective one-body description,
with the local density of states (sampled by the probe) given by $\bar{A}(\omega)$,
and the occupation probabilities given by the local nonequilibrium distribution of the system $f_{s}$:
\begin{equation}
\begin{aligned}
S_{s}\equiv S[f_{s}(\omega)]= -\int_{-\infty}^{\infty}&d\omega\bar{A}(\omega)[f_{s}(\omega)\ln{f_{s}(\omega)}\\&+(1-f_{s}(\omega))\ln{(1-f_{s}(\omega))}].
\label{SystemEntropy}
\end{aligned}
\end{equation}
Within elastic transport theory, it can be shown that $0\leq f_{s}\leq1$ (Appendix\ \ref{appendLocalEntropy})
and therefore $S_{s}$ in Eq.\ (\ref{SystemEntropy}) is real and positive.
$S_{s}$ also correctly reproduces two known limiting cases for the entropy of a system of independent fermions:  (i) in equilibrium, $S_{s}$ gives the 
correct entropy of the subsystem sampled by the probe; (ii) $S_{s}$ gives the correct nonequilibrium entropy for an entire system of
fermions \cite{LandauLifschitz}.
However, we note that the entropies (\ref{SystemEntropy}) of the various subsystems of a quantum system are not additive out of equilibrium. 
The definition of local nonequilibrium entropy given by Eq.\ (\ref{SystemEntropy}) differs from that proposed in Ref.\ \onlinecite{Galperin15}.

We also define the local entropy of the corresponding local equilibrium state of the system if its local spectrum were populated by
the probe's equilibrium distribution function $f_{p}$:
\begin{equation}
\begin{aligned}
S_{p}\equiv S[f_{p}(\omega)]= -\int_{-\infty}^{\infty}&d\omega\bar{A}(\omega)[f_{p}(\omega)\ln{f_{p}(\omega)}\\&+(1-f_{p}(\omega))\ln{(1-f_{p}(\omega))}].
\label{ProbeEntropy}
\end{aligned}
\end{equation}
For sufficiently low probe temperatures,
\begin{equation}
S_p \simeq \frac{\pi^2}{3} \bar{A}(\mu_0) k_B T_p,
\label{eq:Sp_lowTp}
\end{equation}
a standard textbook result.
The maximum entropy principle
implies $S_{p}\geq S_{s}$, since the Fermi-Dirac distribution $f(\omega)=f_{p}(\omega)$ maximizes the local entropy $S[f(\omega)]$ subject to the constraints
imposed by Eqs.\ (\ref{BBOccupancy}) and (\ref{BBEnergy}). 
Clearly, $S_{p}\rightarrow 0$ as $T_p\rightarrow 0$, which implies $S_{s}\rightarrow 0$ as $T_p\rightarrow 0$ (third law of thermodynamics).

We propose the local entropy deficit $\Delta{S}=S_{p}-S_{s}$ as a suitable metric quantifying the
departure from local equilibrium. 
However, it is important to note that the mean local spectrum $\bar{A}(\omega)$
varies significantly from point to point within the nanostructure depending upon the local probe-system coupling (especially in the tunneling regime)
and limits the
use of $\Delta{S}$ while comparing the `distance' from equilibrium for points within the nanostructure. 
The situation is analogous to that of a dilute gas, which can have a very low entropy per unit volume even if it has a very high entropy per particle.
We note that states far from the equilibrium Fermi energy $\mu_{p}$ contribute negligibly to the entropy since $\lim_{f\to0}S[f]=\lim_{f\to1}S[f]=0$, 
and therefore introduce a normalization averaged over the thermal window of the probe:
\begin{equation}
{\cal{N}} 
=\int_{-\infty}^{\infty}d\omega\frac{\bar{A}(\omega)}{\Tr{A(\omega)}}\bigg(\frac{-{\partial{f_{p}}}} {{\partial{\omega}}}\bigg).
\label{normalization}
\end{equation}
We define the local entropy-per-state of the system $s_{s}$ and that of the corresponding local equilibrium distribution $s_p$ as
\begin{align}
s_{s}&=\frac{S_{s}}{\cal{N}},
\label{SystemPerState}\\
s_{p}&=\frac{S_{p}}{\cal{N}}.
\label{ProbePerState}
\end{align}
$\Delta{s}=s_{p}-s_{s}$ quantifies the per-state `distance' from local equilibrium.
We present numerical calculations of the local entropy-per-state in Sec.\ {\ref{sec:results}} and discuss its implications.

\section{Local temperatures near absolute zero}
\label{sec:Tp_near_zero}

Our analyses in this article consider a quantum conductor that is placed in contact with two electron reservoirs:
a cold reservoir $R1$ and a hot reservoir $R2$.
We are interested in the limiting case where reservoir $R1$ is held near absolute zero ($T_{1}\rightarrow{0}$)
while $R2$ is held at finite temperature ($T_{2}=100K$ in our simulations). We assume no electrical bias ($\mu_{1}=\mu_{2}=\mu_{0}$).

Transport in this regime will be dominated by elastic processes,
and occurs within a narrow thermal window close to the Fermi energy of the reservoirs. It should be noted that, although the transport energy window
is small, the problem is essentially outside the scope of linear response theory owing to the large discrepancies in the derivatives of the Fermi functions
of the two reservoirs. Therefore, the problem has to be addressed with the full numerical evaluation of the currents given by Eq.\ (\ref{NumberHeatCurrent}).

For many cases of interest, transport in a nanostructure is largely dominated by elastic processes. This allows us
to utilize the simpler formula analogous to the multiterminal B{\"u}ttiker formula\,\cite{Buttiker86} given by\,\cite{Sivan86} 
\small
\begin{equation}
 I_{p}^{(\nu)}=\frac{1}{h}\sum_{\alpha}\int_{-\infty}^{\infty}d\omega (\omega-\mu_{p})^{\nu}\myT_{p\alpha}(\omega)\big(f_{\alpha}(\omega)-f_{p}(\omega)\big),
\label{NumberHeatCurrent}
\end{equation}
\normalsize
where the transmission function is given by\ \cite{Datta95}
\begin{equation}
\myT_{p\alpha}(\omega)=\Tr{\Gamma^{p}(\omega)G^{r}(\omega)\Gamma^{\alpha}(\omega)G^{a}(\omega)}.
\label{TMatrix}
\end{equation}

A pure thermal bias, such as the one considered in this article, has been shown to lead to temperature oscillations in small molecular junctions\ \cite{Bergfield13demon} and 1-D conductors \cite{Dubi2009,DiVentra09}.
Temperature oscillations have been predicted in quantum coherent conductors such as graphene\ \cite{Bergfield15}, which allow the oscillations to be
tuned (e.g., by suitable gating) such that they can be resolved under existing experimental techniques \cite{Yu2011,Kim2011,Kim2012,Menges2012} of 
SThM. 
More generally, quantum coherent temperature oscillations can be obtained for quantum systems driven out of equilibrium due to external fields \cite{Caso10,Caso2011}
as well as chemical potential \cite{Bergfield14} and temperature bias of the reservoirs. In
practice, the
thermal coupling of the probe to the 
environment sets limitations on the resolution of a scanning thermoelectric probe \cite{Bergfield13demon}.
However, in this article, we ignore the coupling of the probe to the ambient
environment, in order to highlight the theoretical limitations on temperature measurements near absolute zero.

In evaluating the expressions for the currents in Eq.\ (\ref{NumberHeatCurrent}) within elastic transport theory, we encounter integrals of the form
${\int_{-\infty}^{\infty}}d\omega{F(\omega)}\big(f_{2}(\omega)-f_{1}(\omega)\big)$. We use the Sommerfeld series given by
\small
\begin{equation}
 \begin{aligned}
 &\int_{-\infty}^{\infty}d\omega F(\omega)\big(f_{2}(\omega)-f_{1}(\omega)\big)=
 \int_{\mu_{1}}^{\mu_{2}} d\omega F(\omega) \\ 
 &+  2\sum_{k}\Theta(k+1) \big[(k_{B}T_{2})^{k+1}F^{(k)}(\mu_{2})-(k_{B}T_{1})^{k+1}F^{(k)}(\mu_{1})\big],\\
 & \ \ \ \ \ k \in \{1,3,5,...\},
  \end{aligned}
  \label{RiemannSeries}
\end{equation}
\normalsize 
where we use the symbol $\Theta$ that relates to the Riemann-Zeta function as $\Theta(k+1)=\big(1-\frac{1}{2^k}\big)\zeta(k+1)$ and $f_{\alpha}(\omega)$ is the Fermi-Dirac
distribution of reservoir $\alpha$.
The second term on the $r.h.s$ of Eq.\ (\ref{RiemannSeries}) accounts for the exponential tails in $\big(f_{2}(\omega)-f_{1}(\omega)\big)$, and its contribution depends
on the changes to the function $F(\omega)$ in the neighbourhoods of $\omega=\mu_{1}$ and $\omega=\mu_{2}$ and can generally be truncated using a
Taylor series expansion for most well-behaved functions $F(\omega)$. The $l.h.s$ of Eq.\ (\ref{RiemannSeries})
is bounded if $F(\omega)$ grows slower than exponentially for $\omega\rightarrow\pm\infty$ and is satisfied by the current integrals in Eq.\ (\ref{NumberHeatCurrent}).

\subsection{Constant trasmissions}
\label{sec:constant_trans}

In order to make progress analytically, we consider first the case of constant transmissions:
\begin{equation}
\myT_{p\alpha}(\omega)=\myT_{p\alpha}(\mu_{0}) \equiv \myT_{p\alpha}.
\end{equation}
This is a reasonable assumption because the energy window involved in transport is of the order of the thermal energy of the hot reservoir ($k_{B}T_{2}\approx25$meV, at room temperature)
and we may expect no great changes to the transmission function.
In this case the series (\ref{RiemannSeries}) for the number current contains no temperature terms at all,
while the heat current contains terms quadratic in the temperature. It is easy to see that the expression for the number current into the probe becomes
\begin{equation}
  I_{p}^{(0)} = \frac{1}{h}\sum_{\alpha}\myT_{p\alpha}(\mu_{\alpha}-\mu_p),
  \label{ConstantTransmissionNumberCurrent}
\end{equation}
and the heat current into the probe is given by
\begin{equation}
  I_{p}^{(1)} = \frac{1}{h}\sum_{\alpha}\bigg(\myT_{p\alpha}\frac{(\mu_{\alpha}-\mu_p)^{2}}{2} +\frac{\pi^{2}k_{B}^{2}}{6}\myT_{p\alpha}(T_{\alpha}^{2}-T_{p}^{2})\bigg).
  \label{ConstantTransmissionHeatCurrent}
\end{equation}
Eq.\ (\ref{ConstantTransmissionNumberCurrent}) does not depend on the temperature and can be solved readily:
\begin{equation}
\mu_{p}=\mu_{0},
\end{equation}
since $\mu_{1}=\mu_{2}=\mu_{0}$ and Eq.\ (\ref{ConstantTransmissionHeatCurrent}) is solved by
\begin{equation}
T_{p}=\sqrt{\frac{\myT_{p1}T_{1}^{2}+\myT_{p2}T_{2}^{2}}{\myT_{p1}+\myT_{p2}}}.\\
\end{equation}
In this article, we are primarily interested in temperature measurements near absolute zero and work in the limit $T_{1}\rightarrow0$
which yields
\begin{equation}
T_{p}=\sqrt{\frac{\myT_{p2}}{\myT_{p1}+\myT_{p2}}}T_{2}.
\label{eqn1}
\end{equation}
We have $T_{p}\rightarrow{0}$ as $\myT_{p2}\rightarrow{0}$. Indeed, when the system
is decoupled from the hot reservoir $R2$, the probe would read the temperature of reservoir $R1$.

\subsection{Transmission node}
\label{sec:trans_node}

The analysis of the previous section suggests that a suppression in the transmission from the finite-temperature reservoir $R2$ results
in probe temperatures in the vicinity of absolute zero. In quantum coherent conductors, destructive interference gives rise to nodes in the
transmission function. In this section, we consider a case where the transmission from $R2$ into the probe
has a node at the Fermi energy.
In the vicinity of such a node, generically, the transmission probability varies quadratically with energy:
\begin{equation}
\myT_{p2}(\omega)=\frac{1}{2}\myT_{p2}^{(2)}(\omega-\mu_{0})^{2},
\end{equation}
while the transmission from the cold reservoir $R1$ may still be treated as a constant:
\begin{equation}
\myT_{p1}(\omega)=\myT_{p1}.
\label{ConstantProbeTransmission}
\end{equation}
Applying the Sommerfeld series (\ref{RiemannSeries}) for the number current gives us
\begin{equation}
\begin{aligned}
  I_{p}^{(0)} = &\frac{1}{h}\bigg(\myT_{p1}(\mu_{0}-\mu_p)-\frac{\myT_{p2}^{(2)}}{6}(\mu_{p}-\mu_{0})^3\\
   &+ \frac{\pi^{2}}{6}\myT_{p2}^{(2)}(\mu_{0}-\mu_{p})k_{B}^{2}T_{p}^{2}\bigg),
  \end{aligned}
  \label{QNodeNumberCurrent}
\end{equation}
where the $k_{B}T_{2}$ term is still missing since the first derivative of $\myT_{p2}(\omega)$ vanishes at $\mu_{2}=\mu_{0}$.
We note that Eq. (\ref{QNodeNumberCurrent}) admits a single real root at 
\begin{equation}
\mu_{p}=\mu_{0}.
\end{equation}
With this solution, we can write down the equation for the heat current as
\begin{equation}
  I_{p}^{(1)} = \frac{1}{h}\bigg(\frac{\pi^{2}k_{B}^2}{6}\myT_{p1}(T_{1}^{2}-T_{p}^{2})+\frac{7}{8}\frac{\pi^{4}k_{B}^{4}}{15}\myT_{p2}^{(2)}(T_{2}^{4}-T_{p}^{4})\bigg),
  \label{QNodeHeatCurrent}
\end{equation} 
which gives us a simple quadratic equation in $T_{p}^2$.
We note that the above equation is monotonically decreasing in $T_{p}$ for all positive values of temperature. There exists a unique solution to Eq.\ (\ref{QNodeHeatCurrent})
in the interval $T_{1}< T_{p}< T_{2}$, since $I_{p}^{(1)}(T_{p})$ undergoes
a sign change between these two values, and is also the only positive solution due to monotonicity. 
Physically, this solution is reasonable since we expect a temperature measurement to be within the interval $(T_{1},T_{2})$
in the absence of an electrical bias. It is straight-forward to write down the exact solution to Eq.\ (\ref{QNodeHeatCurrent}) (see Appendix\ \ref{append:ExactSolution}). However,
we simplify the expression for $T_{p}$ by noting that
\begin{equation}
\myT_{p2}^{(2)}(k_{B}T_{2})^{2}\ll\myT_{p1},
\end{equation}
that is, the transmission into the probe from $R2$ within a thermal energy window $k_{B}T_{2}$ in the presence of a node,
is small in comparison to the transmission from $R1$. The approximate
solution for $T_{p}$ then becomes
\begin{equation}
T_{p}= \sqrt{\frac{7\pi^{2}}{20}\frac{\myT_{p2}^{(2)}(k_{B}T_{2})^{2}}{\myT_{p1}}}T_{2},\label{eqn2}
\end{equation}
where, as before, we have taken the limiting case where $T_{1}\rightarrow0$.

\subsection{Higher-order destructive interferences}
\label{sec:higher_node}

Although a generic node obtained in quantum coherent transport depends quadratically on the energy,
it is possible to obtain higher-order ``supernodes" in some systems \cite{Bergfield10}. In the vicinity of such a supernode, the
transmission function can be written as
\begin{equation}
\myT_{p2}(\omega)=\frac{1}{2n!}\myT_{p2}^{(2n)}\big|_{\omega=\mu_{0}}(\omega-\mu_{0})^{2n},
\label{node}
\end{equation}
while the transmission from $R1$ may still be approximated by Eq.\ (\ref{ConstantProbeTransmission}). Exact expressions for the currents can again be evaluated
using Eq.\ (\ref{RiemannSeries}). The expression for the number current becomes
\small
\begin{equation}
\begin{aligned}
&I_{p}^{(0)} =\frac{1}{h}\bigg(\myT_{p1}(\mu_{0}-\mu_p)-\frac{\myT_{p2}^{(2n)}}{(2n+1)!}(\mu_{p}-\mu_{0})^{2n+1}\\
+& 2\sum_{k\in odd}\Theta(k+1) \big[(k_{B}T_{2})^{k+1}\myT_{p2}^{(k)}(\mu_{0})-(k_{B}T_{p})^{k+1}\myT_{p2}^{(k)}(\mu_{p})\big]\bigg),
\label{NodeNumberCurrent}
\end{aligned}
\end{equation}
\normalsize
and we have
\begin{equation}
\myT_{p2}^{(k)}(\mu_{0})=0, \ \ \ \forall k\in \{1,3,5,...\}.
\end{equation}
Now, with
\begin{equation}
\mu_{p}=\mu_{0},
\end{equation}
every term on the $r.h.s$ of Eq.\ (\ref{NodeNumberCurrent}) vanishes.
With this solution for $\mu_{p}$, we proceed to write the equation for the heat current. Using Eq.\ (\ref{RiemannSeries}) with \small$F(\omega)=(\omega-\mu_{0})\myT_{p\alpha}(\omega)$
\normalsize, we obtain only one nonvanishing derivative for each reservoir, that is,
\small$F^{(1)}(\mu_{0})= \myT_{p1}(\mu_{0})$\normalsize \ for $R1$ and
\small$F^{(2n+1)}(\mu_{0})= (2n+1)\myT_{p2}^{(2n)}$\normalsize\ for $R2$. Therefore,
\small
\begin{equation}
\begin{aligned}
I_{p}^{(1)} =&\frac{2}{h}\bigg( \Theta(2)\myT_{p1}\big[(k_{B}T_{1})^2-(k_{B}T_{p})^2\big]\\
&+(2n+1) \Theta(2n+2)\myT_{p2}^{(2n)}\big[(k_{B}T_{2})^{2n+2}-(k_{B}T_{p})^{2n+2}\big]\bigg),
\label{NodeHeatCurrent}
\end{aligned}
\end{equation}
\normalsize
which is a polynomial equation in $T_{p}$ of degree (2n+2). We can rewrite Eq.\ (\ref{NodeHeatCurrent}) as
a polynomial $p(x)$ in $x= {T_{p}}/{T_{2}}$:
\begin{equation}
p(x)=x^{2} + \lambda_{n} x^{2n+2} - \lambda_{n} = 0,
\label{poly}
\end{equation}
where we have taken $T_{1}\rightarrow0$ for $R1$, and $\lambda_{n}$ is a dimensionless quantity given by
\begin{equation}
\lambda_{n} = (2n+1)\frac{\Theta(2n+2)}{\Theta(2)}\bigg(\frac{\myT_{p2}^{(2n)}(k_{B}T_{2})^{2n}}{\myT_{p1}}\bigg).
\label{lambda}
\end{equation}
We will have $\lambda_{n}\ll1$ for a suitable energy window set by $k_{B}T_{2}$, since the transmission into the probe from $R2$ suffers
destructive interference at the Fermi energy ($\mu_{0}$). If the thermal energy is large enough, then this approximation may no longer hold. In any case,
it is possible to define a temperature $T_{2}$ so that this approximation is strongly valid. Under the validity of this approximation, 
the solution to Eq.\ (\ref{poly}) can be written using perturbation theory as
\begin{equation}
x= \sqrt{\lambda_{n}}\bigg(1+ \gcal{O}{\big(\lambda^{n+1}_{n}\big)}\bigg),
\label{polysolution}
\end{equation}
with corrections that are of much higher-order in $\lambda_{n}$.
The solution
for $T_{p}$ given in Eq.\ (\ref{polysolution}) reduces to Eq.\ (\ref{eqn1}) for the case with constant transmissions by setting $n=0$,
and to the approximate result obtained in Eq.\ (\ref{eqn2}) in the presence of a node by setting $n=1$.
We note that higher-order interference effects cause the probe temperature to decay more rapidly with respect to $T_{2}$ since $T_{p} \sim T_{2}^{n+1}$, that is,
when $T_{2}$ is halfed, $T_{p}$ is reduced by a factor of $2^{n+1}$. Now, if we consider the limiting case
where $R2$ is also cooled to absolute zero, $T_{2}\rightarrow0$, Eq.\ (\ref{polysolution}) implies that $T_{p}\rightarrow0$ at
least as quickly as $T_{2}$ (in the absence of destructive interference, i.e., $n=0$) or quicker (when there is destructive interference, i.e., $n\geq1$).

It should be noted that the polynomial given in Eq.\ (\ref{poly}) is monotonically increasing for all positive $x$ and furthermore,
there is only one positive root since $p(0)<0$ and $p(1)>0$. Stated in terms of $T_{p}$, this implies the existence of a unique solution
for the measured temperature $T_{p}$ in the interval $T_{1}<T_{p}<T_{2}$, as noted in the previous subsection ($T_{1}$ has been set to zero in Eq.\ (\ref{poly})).
The sign of $p(x)$ essentially tells us the direction of heat flow for a temperature bias of the probe
with respect to its equilibrium value. $p(x)<0$\ ($p(x)>0$) corresponds to heat flowing into\ (out of) the probe, since we changed the sign in
re-arranging Eq.\ (\ref{NodeHeatCurrent}) to Eq.\ (\ref{poly}).
The monotonicity of $p(x)$ is therefore equivalent to the Clausius statement of the second law of thermodynamics,
and also ensures the uniqueness of the temperature measurement.

\section{Numerical results and discussion}
\label{sec:results}

In order to illustrate the above theoretical results, and further characterize the local properties of the nonequilibrium steady state,
we now present numerical calculations for several molecular junctions with $\pi$-conjugation.
In all of the simulations, the molecule is connected to a cold reservoir $R1$ at $T_1=0$K and a hot reservoir $R2$ at $T_2=100$K.
There is no electrical bias; both electrodes have chemical potential $\mu_0$.
The temperature probe is modeled as an atomically-sharp Au tip scanned horizontally at a constant vertical height of 
3.5\AA\ above the plane of the carbon nuclei in the molecule (tunneling regime). 

The molecular Hamiltonian is described within 
H{\"u}ckel theory,
$H_{\rm mol}= \sum\limits_{<i,j>}t_{ij}d_{i}^{\dagger}d_{j} +{\rm h.c}$, 
with nearest-neighbor hopping matrix element $t=-2.7$eV.
The coupling of the molecule with the reservoirs is described by the tunneling-width matrices $\Gamma^{\alpha}$.
The retarded Green's function of the junction is given by $G^{r}(\omega)=[\mathbb{S}\omega-H_{\rm mol}-\Sigma_{T}(\omega)]^{-1}$, where 
$\Sigma_{T}=-i\sum_{\alpha}\Gamma^{\alpha}/2$ is the tunneling self-energy.
We take the lead-molecule couplings in the broad-band limit, i.e., $\Gamma^{\alpha}_{nm}(\omega)=\Gamma^{\alpha}_{nm}(\mu_{0})$ where
$\mu_{0}$ is the Fermi energy of the metal leads. We also take the lead-molecule couplings to be diagonal matrices
$\Gamma^{\alpha}_{nm}(\omega)=\Gamma_{\alpha}\delta_{nl}\delta_{ml}$ coupled to a single $\pi$-orbital $l$ of the molecule. 
$\mathbb{S}$ is the overlap-matrix between the atomic orbitals on different sites and we take $\mathbb{S}=\mathbb{I}$, 
i.e., an orthonormal set of atomic orbitals. The lead-molecule couplings are taken to be symmetric, with
$\Gamma_{1}=\Gamma_{2}=0.5$eV. 
The non-zero elements of the
system-reservoir couplings for $R1$ (cold) and $R2$ (hot) are indicated with a blue and red square, respectively, corresponding
to the carbon atoms in the molecule covalently bonded to the reservoirs.
The tunneling-width matrix $\Gamma^p$ describing probe-sample coupling is also treated in the broad-band limit
(see Appendix \ref{TipSampleCoupling} for details of the modeling of probe-system coupling).
The probe is in the tunneling regime
and the probe-system coupling is weak (few meV) in comparison to the system-reservoir couplings ($\Gamma_{1}=\Gamma_{2}=0.5$eV). 

\begin{figure*}[tb]
	\centering
	\captionsetup{justification=raggedright,
singlelinecheck=false
}
		\includegraphics[width=3in]{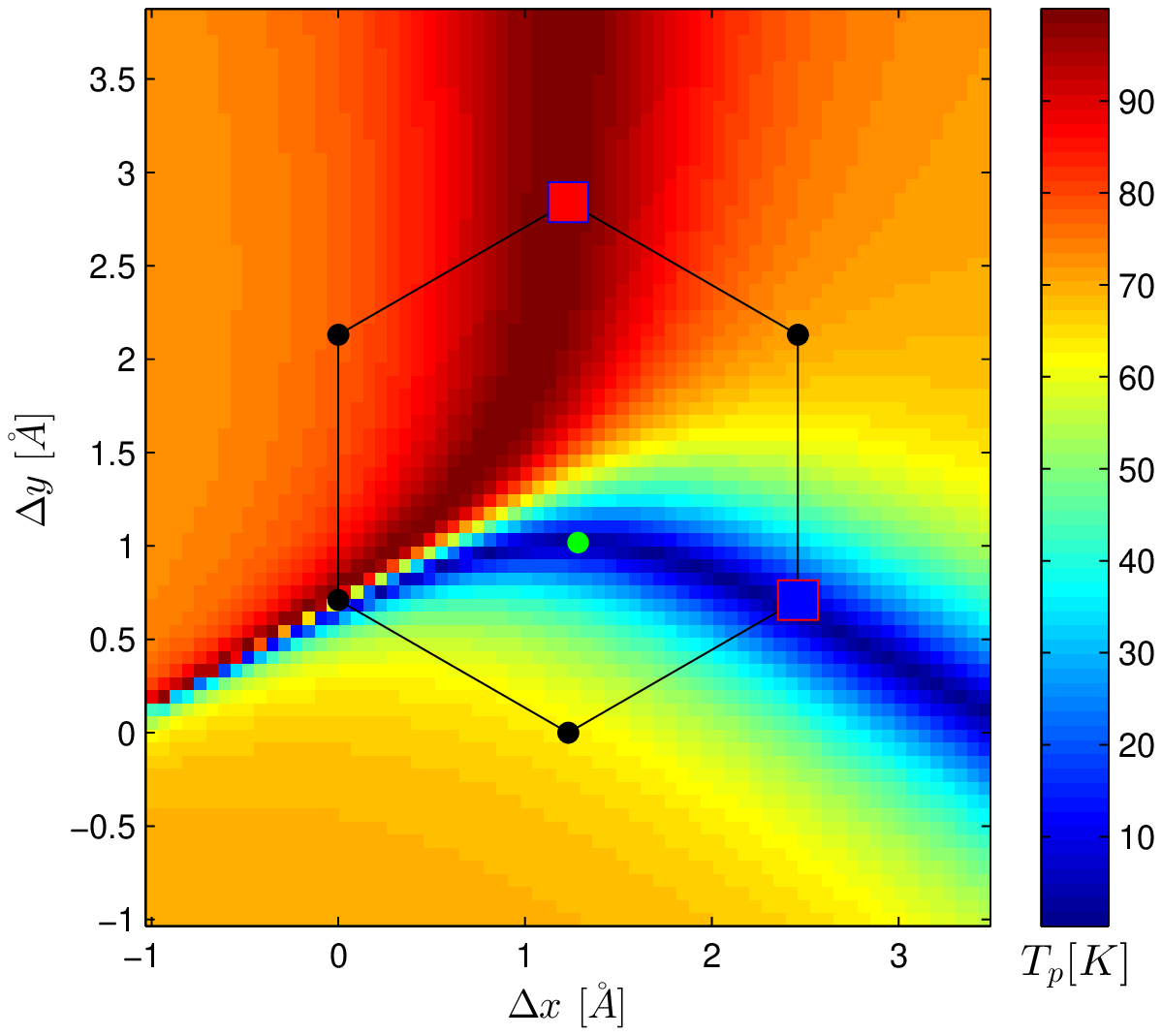}
		\includegraphics[width=3in]{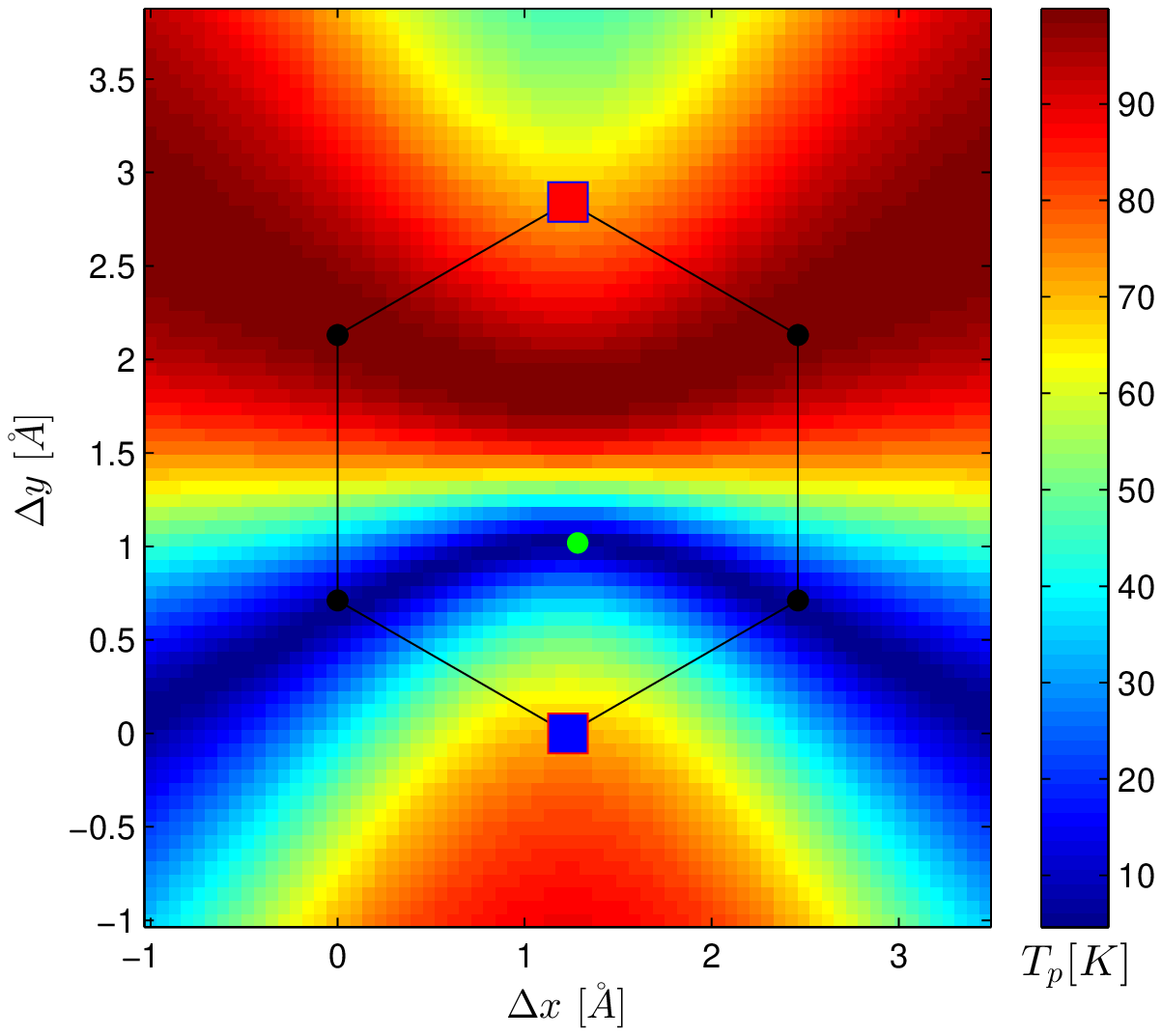}
		\includegraphics[width=3.5in]{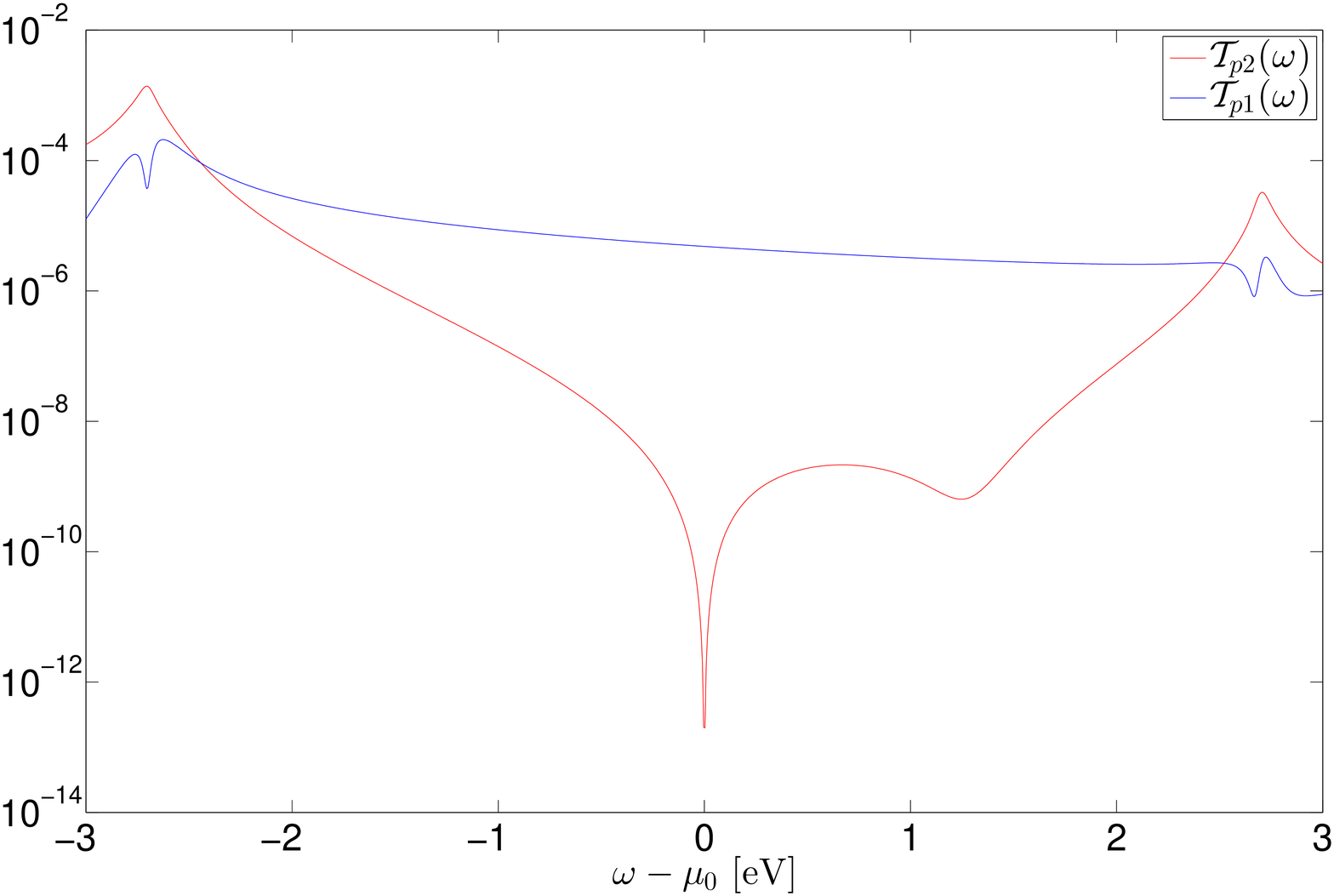}
		\includegraphics[width=3.5in]{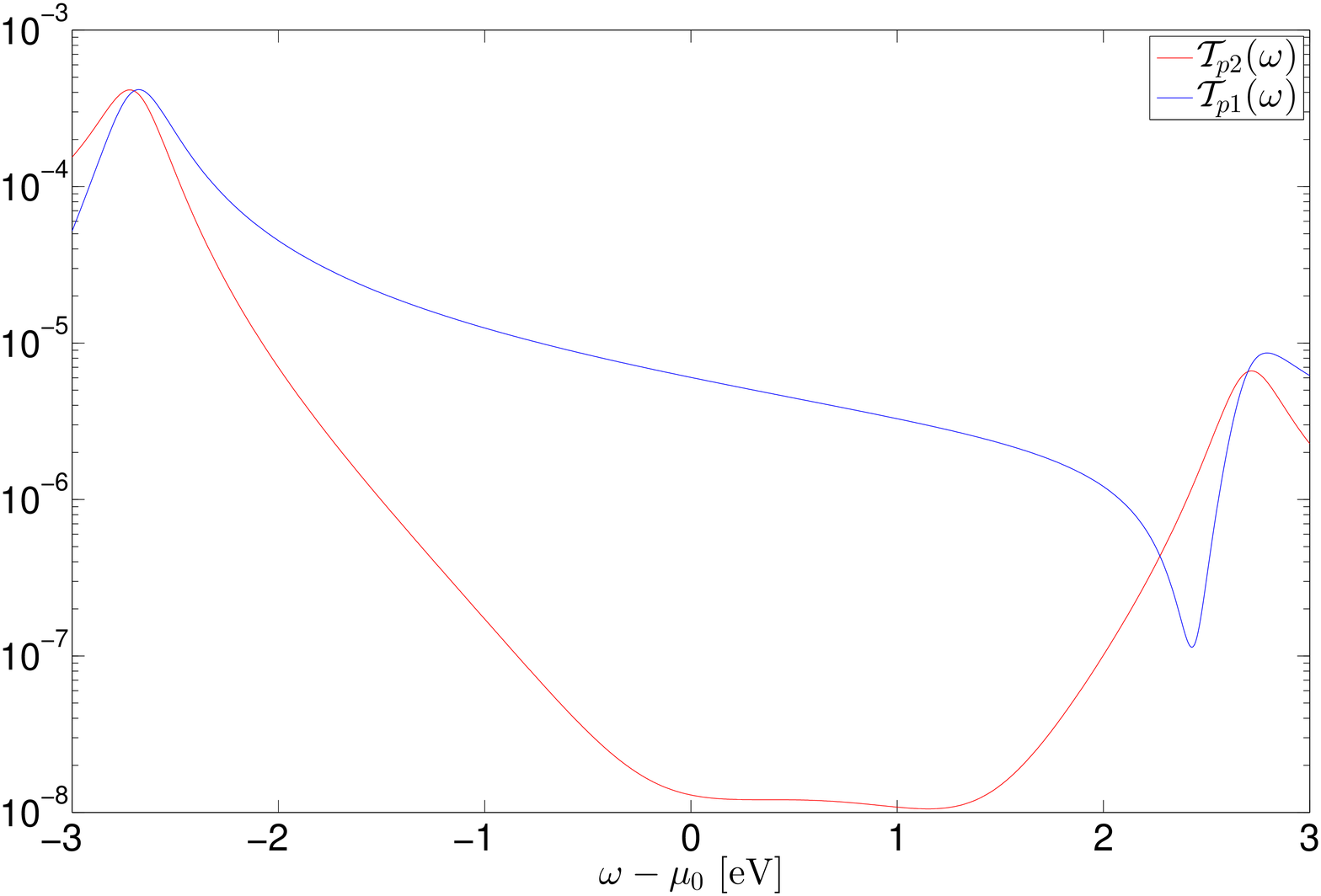}
		\caption{Upper panels: Local temperature distributions for Au-benzene-Au junctions in meta and para configurations, respectively. 
The thermal bias is supplied by cold ($T_{1}=0$K) and hot ($T_{2}=100$K) reservoirs covalently bonded to the atoms indicated by the blue and red squares, 
respectively, and there is no electrical bias.
The probe is scanned at a height of 3.5\AA\ above the plane of the carbon nuclei in the molecule. 
		The green dots shown in the temperature distributions correspond to the coldest temperature found in each of the junction configurations.
		Bottom panels: Transmission probabilities into the probe from $R1$ (cold, i.e., blue curve) and $R2$ (hot, i.e., red curve),
		when the probe is positioned over the coldest spot (shown by the corresponding green dot in the upper panel).
		The existence of a transmission
node in the meta configuration leads to a greatly suppressed probe temperature (see Table\ \ref{table1} for comparison).
		Note the very different vertical scales in the bottom panels.}
	\label{benzene}
\end{figure*}

It must be emphasized that, although we take a noninteracting Hamiltonian for the isolated molecule, our results depend only upon the
existence of transmission nodes, 
which are a characteristic feature of coherent transport, and do {\em not} depend on the particular form of the junction Hamiltonian. 

\subsection{Local temperatures}
\label{sec:results_Tp}

\begin{figure*}[tbh]
	\centering
	\captionsetup{justification=raggedright,
singlelinecheck=false
}
		\includegraphics[width=3.5in]{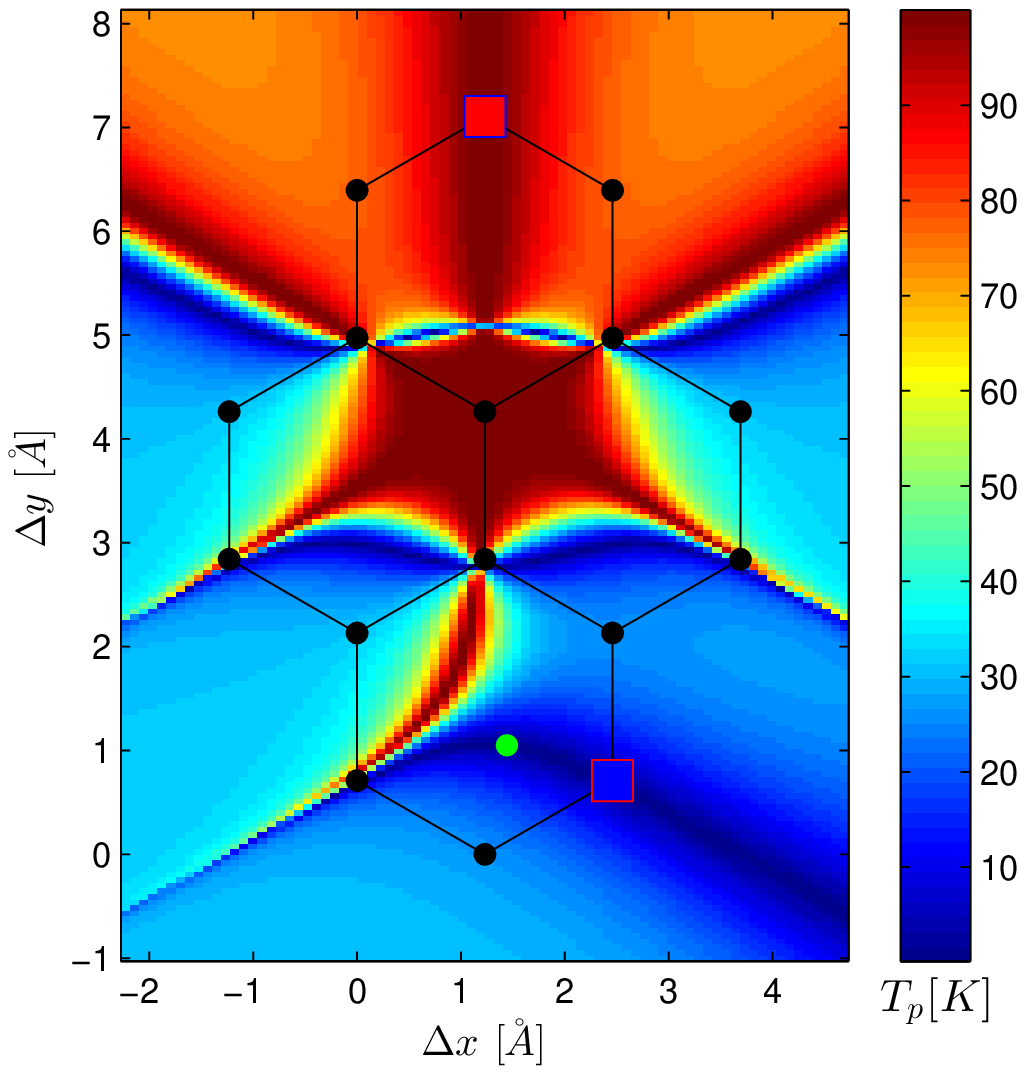}
		\includegraphics[width=3.5in]{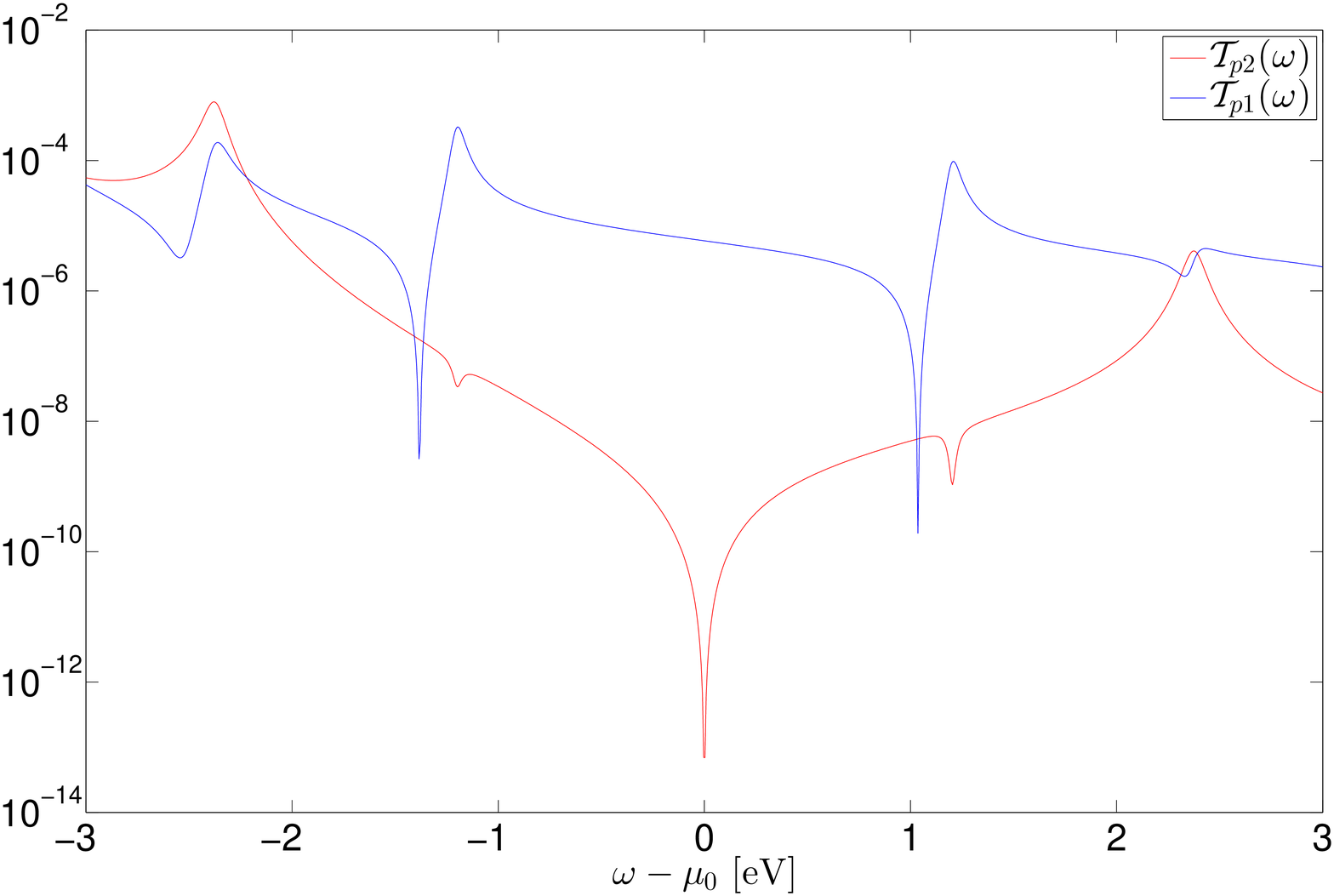}
		\caption{Left panel: Probe temperature distribution in a Au-pyrene-Au junction under the same conditions described in
Fig.\ \ref{benzene}.
The green dot corresponds to the coldest temperature
		found by the search algorithm.
		Right panel: Transmissions into the probe from the hot reservoir $R2$ (red) and the cold reservoir $R1$ (blue) at the coldest
		position, indicated by the green dot on the left. 
		The probe transmission from $R2$ exhibits a (mid-gap) node at the Fermi energy $\mu_{0}$ of the reservoirs, thereby suppressing the temperature
		measured by the probe. 
}
	\label{pyrene}
\end{figure*}

We considered several different molecules and electrode configurations, with and without transmission nodes, 
and searched for the coldest spot in each system (indicated by a green dot in the figures) as measured by the scanning thermoelectric probe.

The local temperature depends on the transmissions from the reservoirs into the probe (Eq.\ (\ref{TMatrix})), determined by 
the local probe-system coupling $\Gamma^{p}$ (see appendix \ref{TipSampleCoupling}), and is thus a function of probe position.
The coldest spot 
was found using a particle swarm optimization technique that minimizes the ratio of the transmissions
to the probe (within a thermal window) from the hot reservoir $R2$ to that of the cold reservoir $R1$, within a search space 
that spans the $z$-plane at 3.5\AA\ and restricted in
the $xy$ direction within 1\AA\ from the edge of the molecule
\footnote{It is necessary to minimize the ratio of transmissions to
find the temperature minima since it is computationally prohibitive to calculate the temperatures at various
points in the search space, within each iteration of the optimization algorithm}.
The numerical solution to Eq.\ (\ref{equilibrium}) was found
using Newton's method. While the algorithm 
was found to converge rapidly for most points (less than 15 iterations), it is still computationally intensive since the evaluation of the currents
given by Eq.\ (\ref{NumberHeatCurrent}) must have sufficient numerical accuracy.
We also note that the minimum probe temperature obtained for each junction does not depend strongly on the
distance between the plane of the scanning probe and that of the molecule. This is explained as follows: the probe temperature must depend upon the relative magnitudes
of the transmissions into the probe from the two reservoirs, and not their actual values. Therefore, the temperature remains roughly independent of the coupling strength
$\Tr{\Gamma^{p}}$.
It has also been previously noted in Ref.\ \onlinecite{Meair14}, that the local temperature measurement showed little change with the coupling strength
even when varied over several orders of magnitude. The restrictions placed on our search space within the optimization algorithm are therefore well justified.

\begin{figure*}[tbh]
  	\centering
  	\captionsetup{justification=raggedright,
singlelinecheck=false
}
 		\includegraphics[width=3.5in]{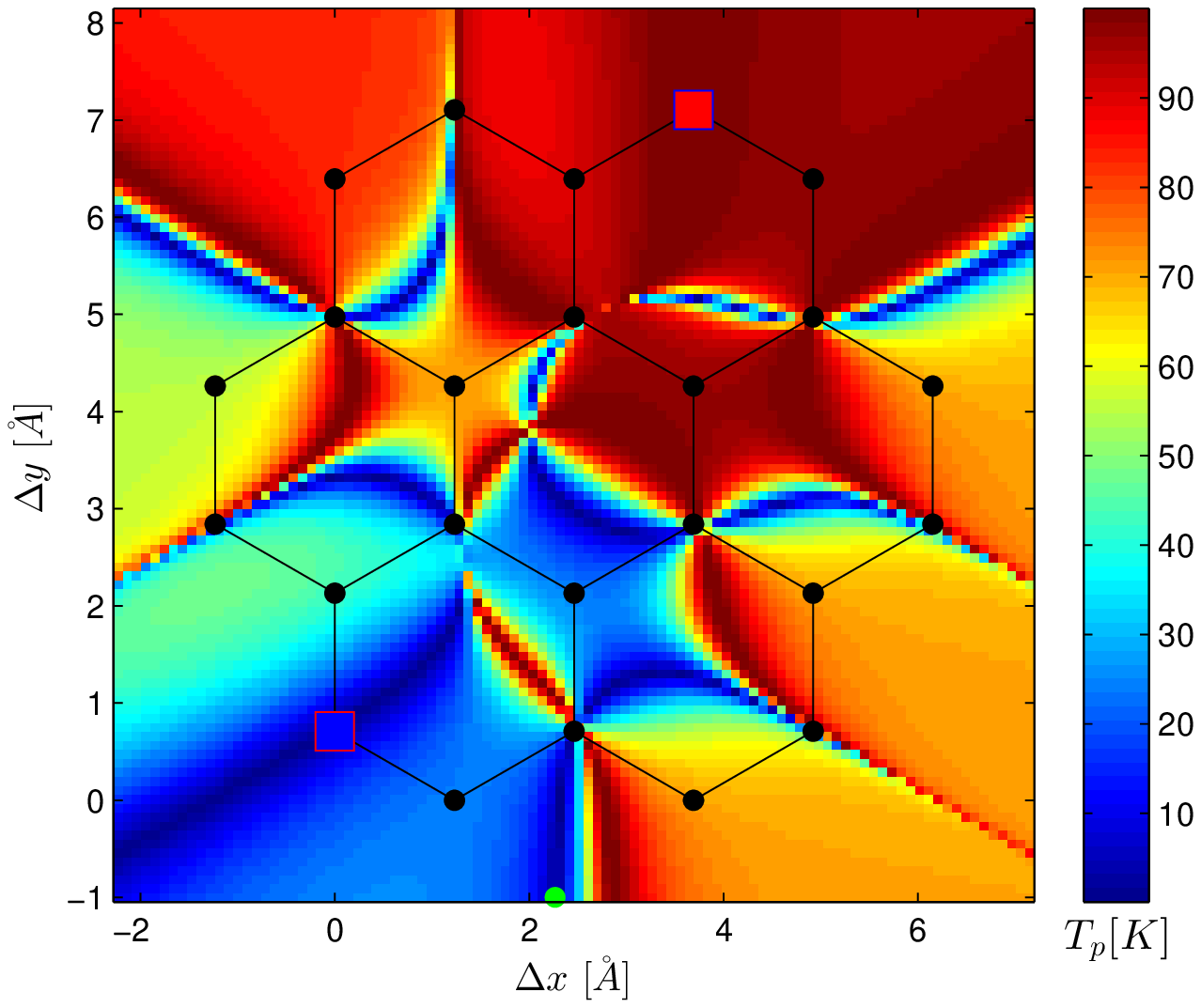}
 		\includegraphics[width=3.5in]{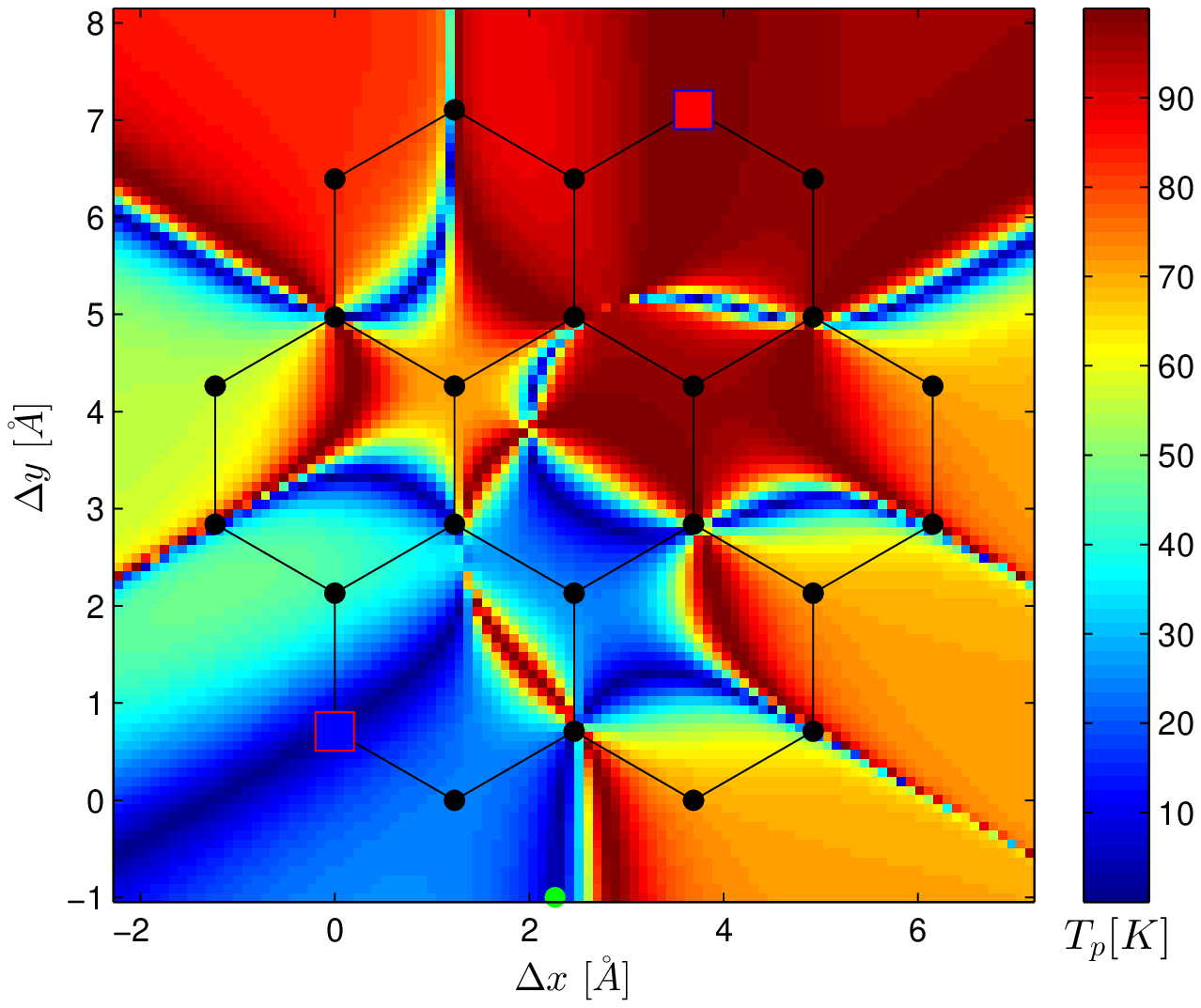}
 		\includegraphics[width=3.5in]{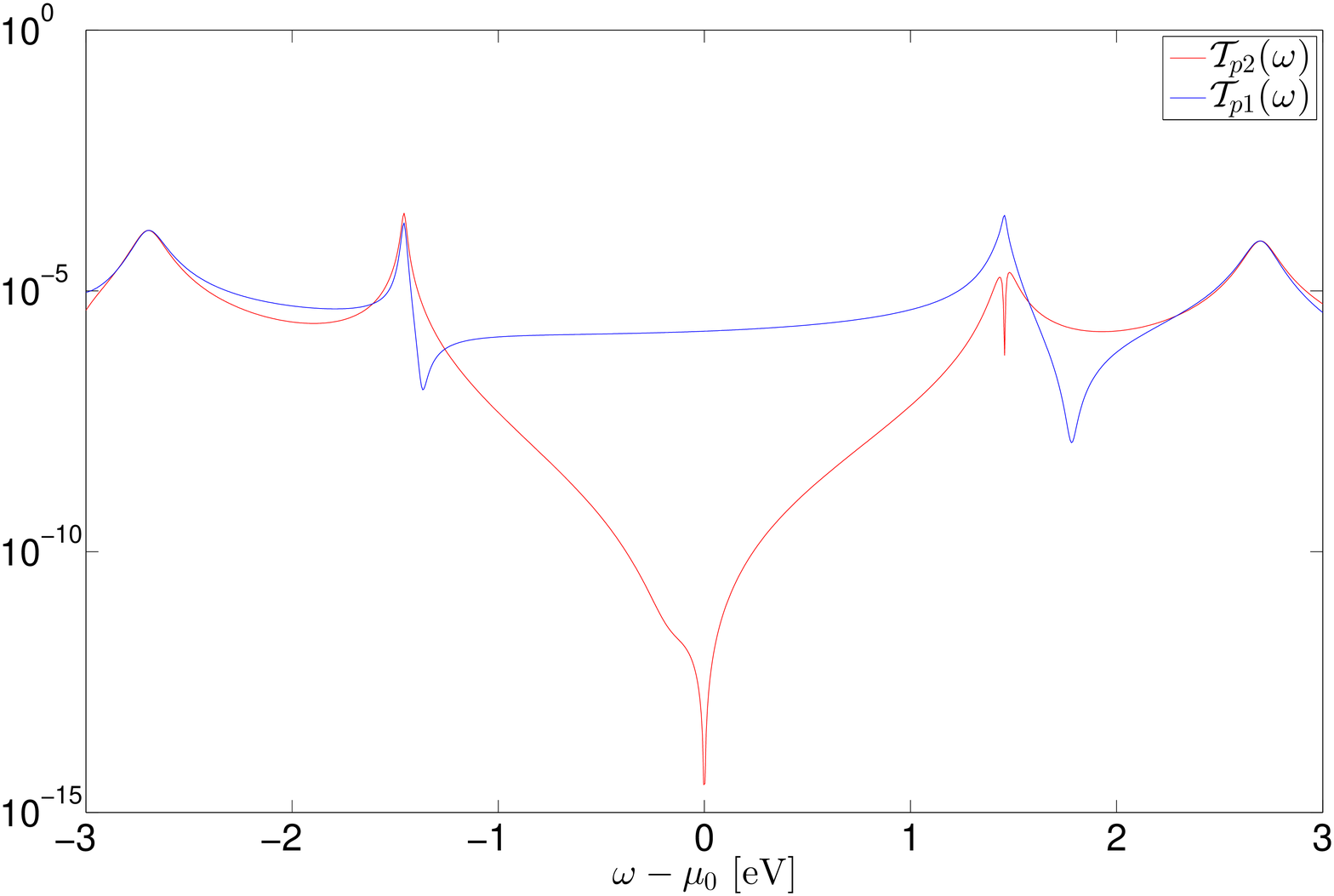}
 		\caption{Upper panels: Probe temperature distributions for a Au-coronene-Au junction
under the same conditions described in Fig.\ \ref{benzene}.
The numerically calculated temperature is on the left, 
and to the right is the analytically calculated temperature using Eq.\ (\ref{eqn1}). 
Although it is in excellent qualitative agreement (and quantitative agreement for the most part), Eq.\ (\ref{eqn1}) poorly estimates the temperature
for the coldest spot (shown in green) due to the existence of a transmission node. Eq.\ (\ref{eqn2}) gives the correct estimation in the presence of a node,
while Eq.\ (\ref{eqn1}) incorrectly predicts $T_{p}=0K$.
The lower panel shows the probe transmissions from the two reservoirs ($R1$ with $T_{1}=0K$ in
blue and $R2$ with $T_{2}=100K$ in red) corresponding to the probe positioned over the coldest spot (shown in
green).}
 	\label{coronene}
\end{figure*}

Fig.\ \ref{benzene} shows the temperature distribution for two configurations of Au-benzene-Au junctions with the chemical potentials
of the metal leads at the middle of the \HOLU gap. The 
mid-gap region is advantageous, since (i) the molecule is charge neutral when the lead chemical potentials are tuned to
the mid-gap energy, and 
(ii) the mismatch between the metal leads' Fermi energy and the mid-gap energy is typically small (less than 1-2 eV for most metal-molecule junctions)
and available gating techniques \cite{Song2009} would be sufficient to tune across the gap. 
In both junctions, the region of lowest temperature passes through the two sites in meta orientation relative to the hot electrode, because the transmission
probability from the hot electrode into the probe is minimum when it is at these locations \cite{Bergfield13demon}.
The meta junction exhibits much lower minimum temperature measurements
since there is a transmission node from the hot reservoir $R2$ at the mid-gap energy, but such nodes are absent in the para junction.
Table\ \ref{table1} shows the coldest temperature found in each of the junctions presented here.

\begin{table}[bt]
  	\captionsetup{justification=raggedright,
singlelinecheck=false
}
\begin{tabular}{|c|c|c|c|}
\hline
Junction       & \begin{tabular}[c]{@{}c@{}}$T_{p}[K]$\\  Numerical\end{tabular} & \begin{tabular}[c]{@{}c@{}}$T_{p}[K]$\\  Analytic\end{tabular} & \begin{tabular}[c]{@{}c@{}}Eq.\end{tabular} \\ \hline
benzene (meta) & 0.154  & 0.1526 & (\ref{eqn2}) \\ \hline
benzene (para) & 4.624   & 4.627 & (\ref{eqn1}) \\ \hline
pyrene         & 0.0821 & 0.0817  & (\ref{eqn2}) \\ \hline
coronene       & 0.0349 & 0.0355 & (\ref{eqn2})  \\ \hline
\end{tabular}
\caption{The table shows the lowest temperatures found in the different junctions considered. All junctions have the same bias conditions: $T_{1}=0K$, $T_{2}=100K$ and no electrical bias. 
The right-most column shows the equation used to compute the temperature analytically. We obtain excellent agreement between the numerical and analytic results.
The para configuration of the benzene junction does not display a node in the
probe transmission spectrum and therefore the minimum probe temperature is not strongly suppressed.}
\label{table1}
\end{table}

We only present two junction geometries for benzene here to illustrate that the existence of a node in the probe transmission spectrum, at the mid-gap energy,
depends upon the junction geometry. The nodes are also absent in the ortho configuration of benzene, and the lowest temperature found in this case
is similar to that found in the para configuration.

Fig.\ \ref{pyrene} shows the temperature distribution in a gated Au-pyrene-Au junction, and the transmissions into the probe from the two reservoirs at the 
coldest spot.  In general, nodes in the transmission spectrum occur only in a few of the possible junction geometries. 
As in the benzene junctions, the coldest regions in the pyrene junction pass through the sites to which electron transfer from the hot 
electrode is blocked by the {\em rules of covalence} \cite{Bergfield13demon} describing bonding in $\pi$-conjugated systems. 
We note that the temperature distribution shown in Fig.\ \ref{pyrene} differs significantly from that shown in Ref.\ \onlinecite{Bergfield13demon} for
four important reasons: (i) the junction configuration in Fig.\ \ref{pyrene} is asymmetric, while that considered in Ref.\ \onlinecite{Bergfield13demon}
was symmetric; (ii) the thermal coupling $\kappa_{p0}$ of the temperature probe to the ambient environment has been set to zero in
Fig.\ \ref{pyrene} to allow for resolution of temperatures very close to absolute zero, while the probe in Ref.\ \onlinecite{Bergfield13demon}
was taken to have $\kappa_{p0}=10^{-4}\kappa_{p0}$, where $\kappa_{p0} = (\pi^2/3)(k_B^2 T/h) = 2.84 \times 10^{-10}$W/K at $T=300$K is
the thermal conductance quantum \cite{Rego98,Rego99}; (iii) the transport in Fig.\ \ref{pyrene} is assumed to 
take place at the mid-gap energy due to appropriate gating of the junction, while Ref.\ \onlinecite{Bergfield13demon} considered
a junction without gating; and (iv) Ref.\ \onlinecite{Bergfield13demon} considered temperature measurements only in the linear
response regime, while the thermal bias applied in Fig.\ \ref{pyrene} is essentially outside the scope of linear response.
Points (ii)--(iv) also differentiate the results for benzene junctions shown in Fig.\ \ref{benzene} from the linear-response results of
Ref.\ \onlinecite{Bergfield13demon}.

Fig.\ \ref{coronene} shows the temperature distribution in a gated Au-coronene-Au junction exhibiting a node in the probe transmission spectrum.  The junction
shown was one of three such geometries to exhibit nodes (10 distinct junction geometries were considered).
Again, the coldest regions in the junction pass through the sites to which electron transfer from the hot
electrode is blocked by the rules of covalence.
The coronene junction in Fig.\ \ref{coronene}
displays the lowest temperature amongst all the different junctions considered, with a minimum temperature of $T_p=35$mK. 
It should be noted that this temperature would be suppressed by
a factor of 100, i.e., $T_p=350{\mu}$K if $R2$ were held at 10K due to the quadratic scaling of $T_{p}$ with respect to the temperature $T_{2}$ 
[cf.\ Eq.\ (\ref{eqn2})].
Higher-order nodes would produce even greater suppression.

\begin{figure*}[tbh]
\centering
\captionsetup{justification=raggedright,
singlelinecheck=false
}
\hspace{-0.7in}
 		\includegraphics[width=2.9in]{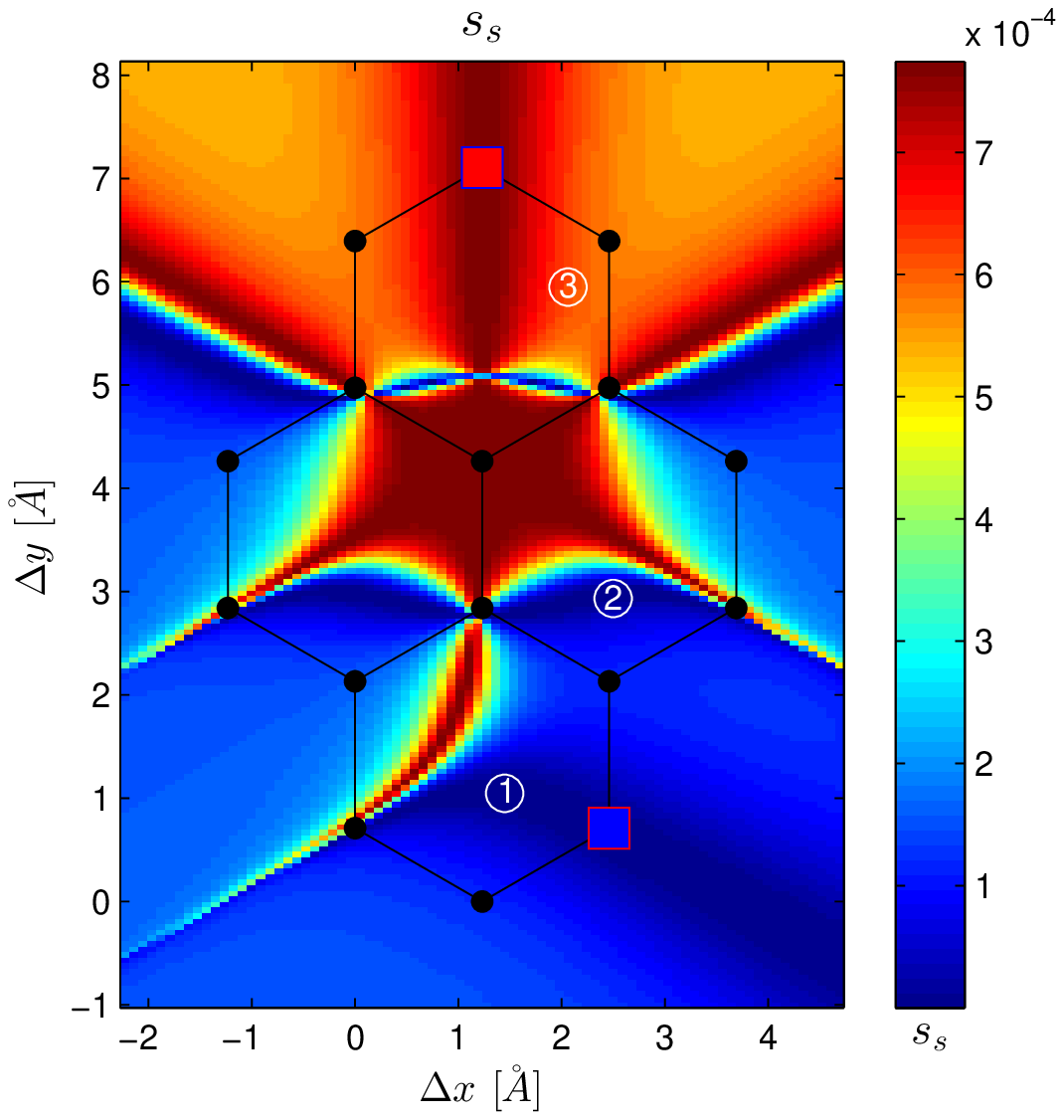}\hspace{-0.53in}
 		\includegraphics[width=2.9in]{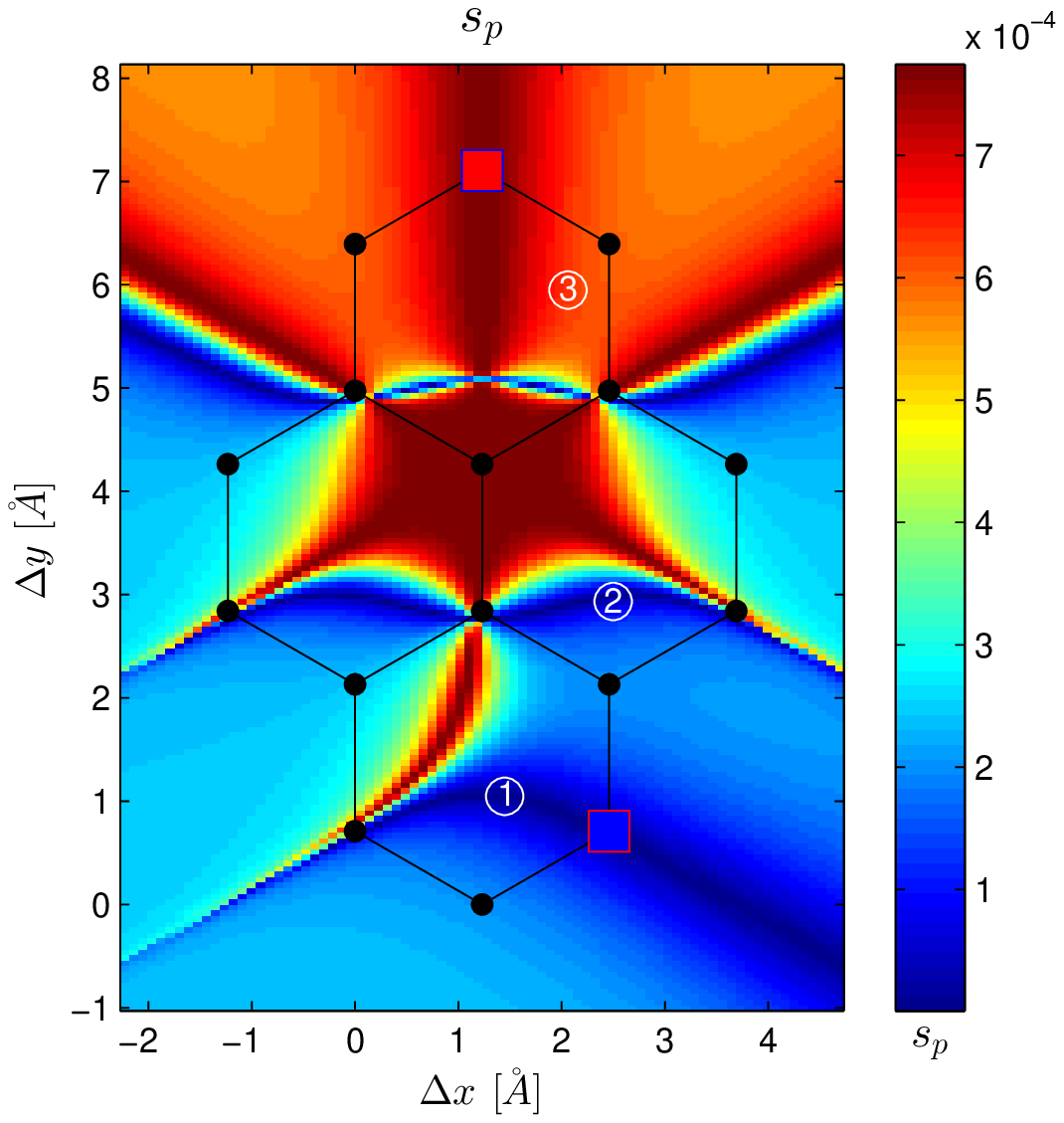}\hspace{-0.53in}
 		\includegraphics[width=2.9in]{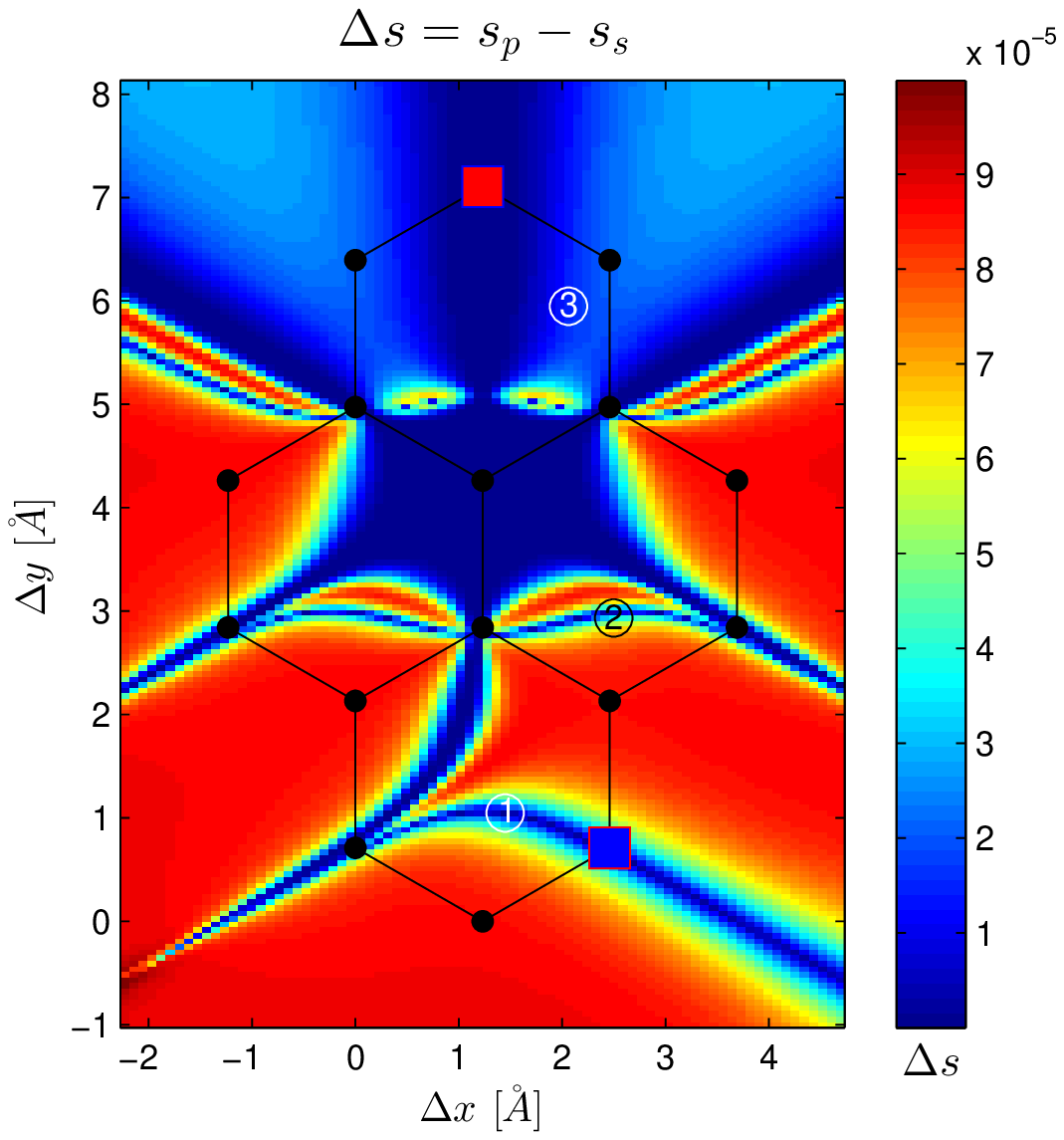} \hspace{-0.6in}\\ \hspace{-0.7in}
 		\includegraphics[width=2.2in]{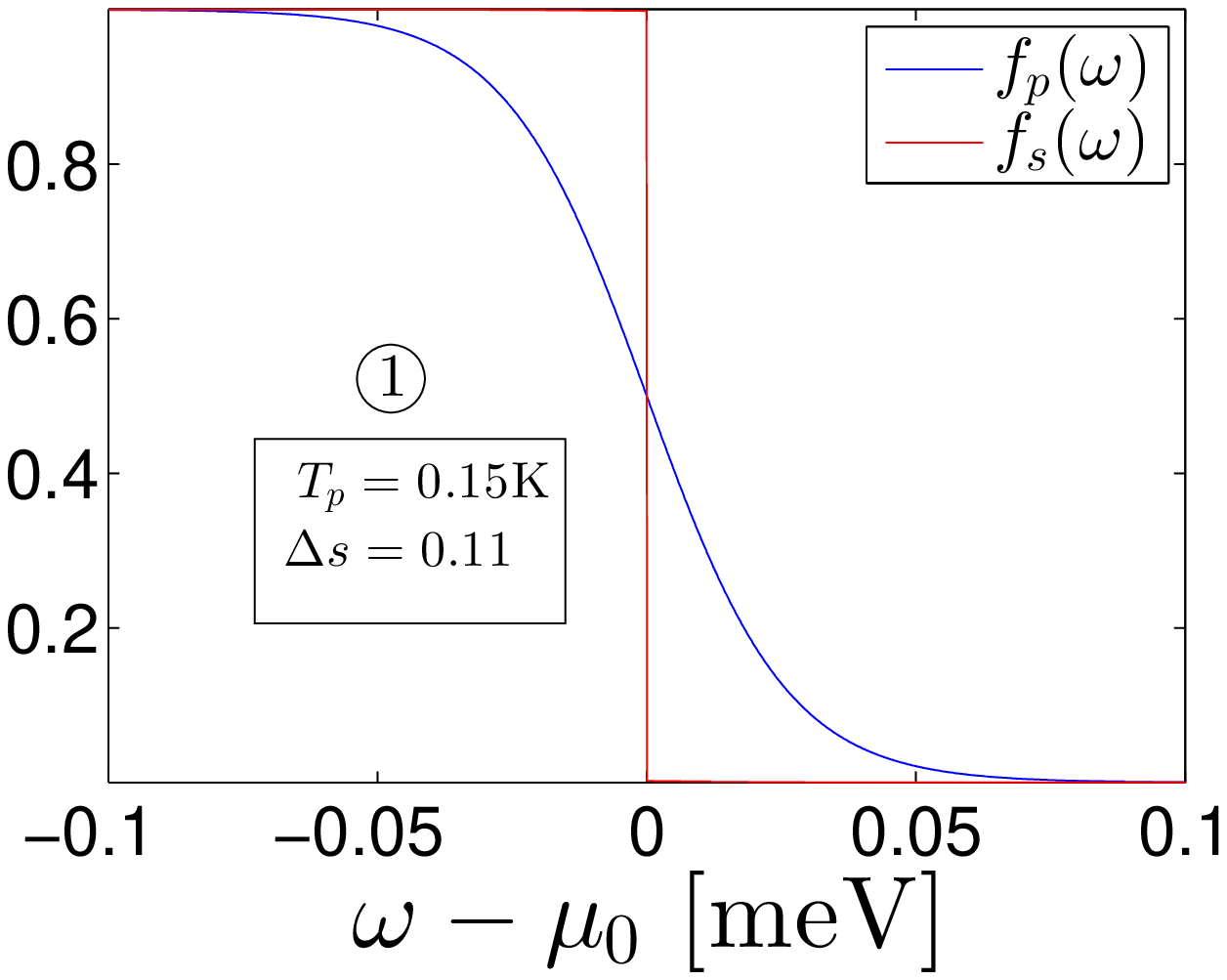} 
 		\includegraphics[width=2.2in]{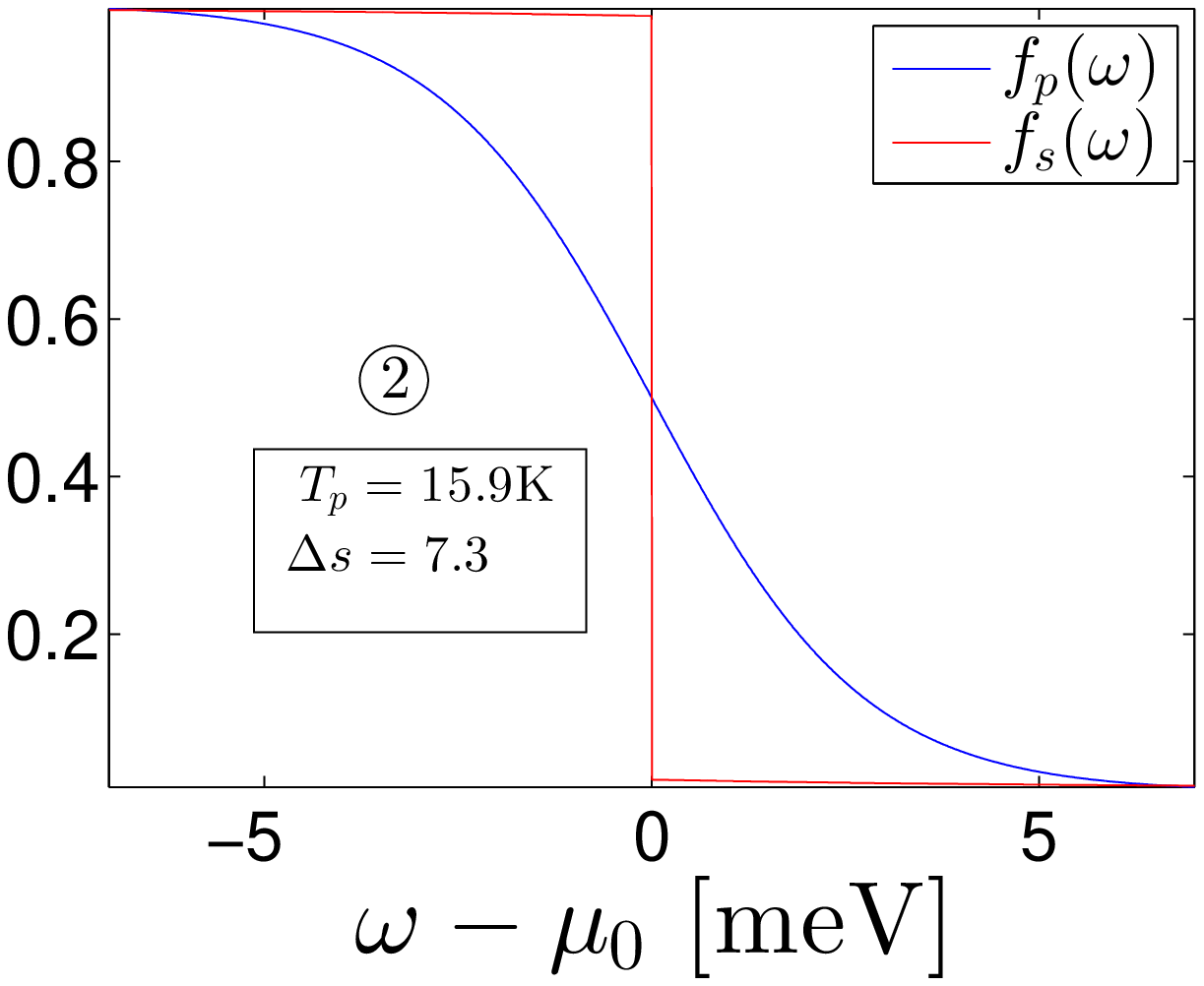} 
 		\includegraphics[width=2.2in]{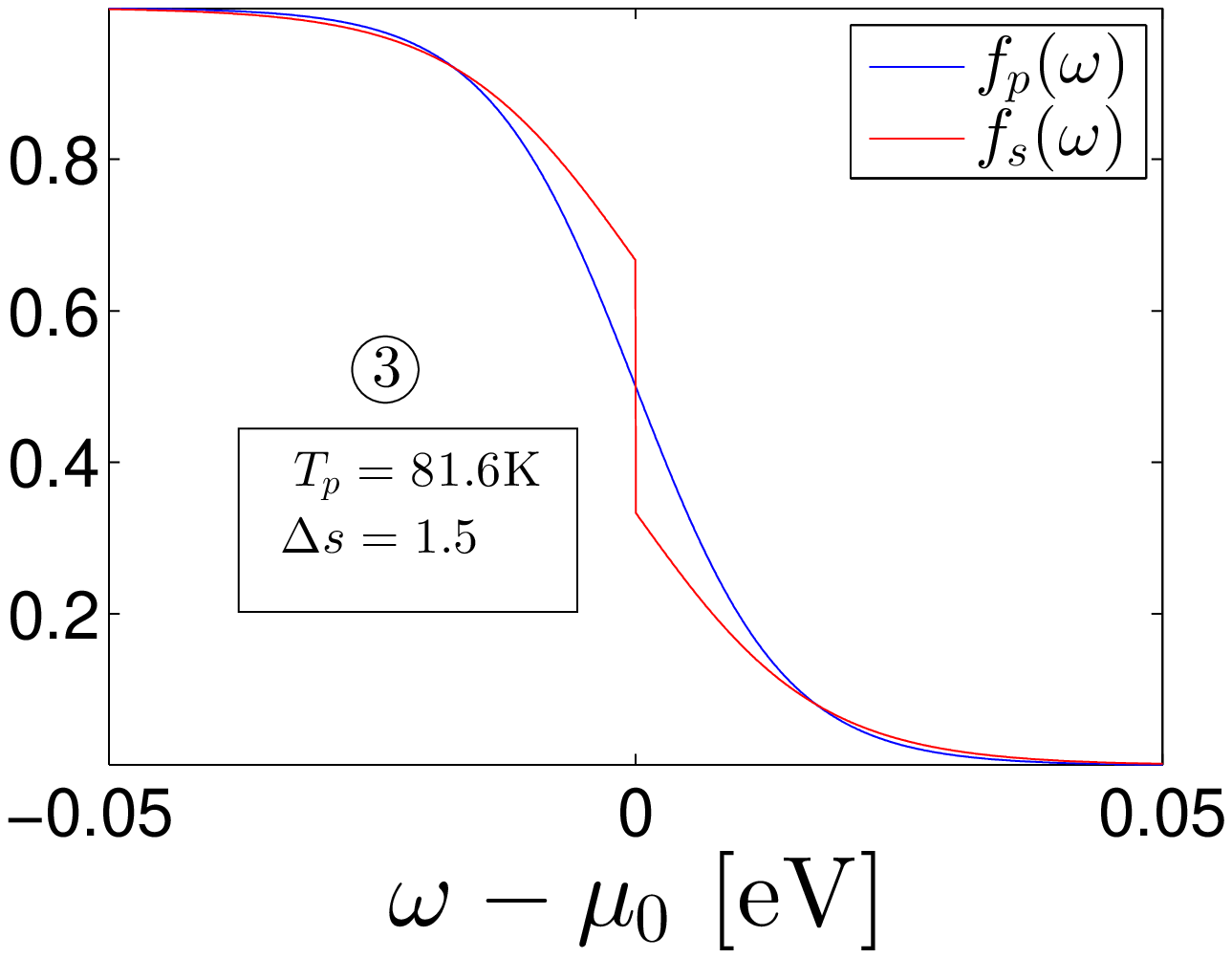} 
\caption{
Upper panels (left to right): The local entropy-per-state of the system $s_{s}$, of the corresponding local equilibrium distribution
$s_{p}$, and the local entropy deficit $\Delta{s}\equiv s_p-s_s$. 
The temperature distribution
for the same junction (with identical bias conditions and sampling of probe positions) is shown in Fig.\ \ref{pyrene},
and we note that it resembles almost exactly the distribution $s_{p}$. 
Lower panels: The distributions $f_{s}$ and $f_{p}$ for three points
shown in the upper panels, each having different probe temperatures $T_{p}=0.15$K, 15.9K and 81.6K, respectively. 
The corresponding entropy deficits are $\Delta{s}=0.11$, 7.3, and 1.5, respectively,  $\times 10^{-5}$.
Point 2, although closer to 0K than point 3 is to 100K, is further from local equilibrium.
} 
\label{entropy}
\end{figure*}

\subsection{Local entropies}
\label{sec:results_S}

The local temperature distributions shown in Figs.\ \ref{benzene}--\ref{coronene} are essentially outside the scope of linear response theory
\cite{Bergfield13demon} since the cold reservoir $R1$ is held at $T_1=0$K, and derivatives of the Fermi function are singular at $T=0$.  However, it is
an open question {\em how far out of equilibrium} these systems are and which regions therein
manifest the most fundamentally nonequilibrium character.  To address such questions quantitatively, we use the concept of local entropy per state
introduced in Sec.\ \ref{sec:entropy}.  In particular, the normalized local entropy deficit $\Delta s\equiv s_p- s_s$ defined through
Eqs.\ (\ref{SystemEntropy})--(\ref{ProbePerState}) allows us to quantify how far the system is from local equilibrium.

Fig.\ \ref{entropy} shows the local entropy distribution of the system $s_{s}$ and that of the corresponding local equilibrium
distribution $s_{p}$, defined by Eqs.\ (\ref{SystemPerState}) and (\ref{ProbePerState}), respectively,
for the Au-pyrene-Au junction considered earlier in Fig.\ \ref{pyrene}. 
The $s_p$ distribution strongly resembles the temperature distribution shown in Fig.~\ref{pyrene}, consistent with the fact (\ref{eq:Sp_lowTp})
that the equilibrium
entropy of a system of fermions is proportional to temperature at low temperatures.  This resemblence is only manifest in the properly-normalized
entropy per state $s_p$; the spatial variations of $S_p$ are much larger, and stem from the orders-of-magnitude variations of the local density of
states $\bar{A}(\mu_0)$.  The nonequilibrium entropy distribution $s_s$ of the system qualitatively resembles $s_p$, but everywhere satisfies the 
inequality $s_s\leq s_p$ (see Sec.\ \ref{sec:entropy}).  $s_s\rightarrow 0$ whenever $T_p\rightarrow 0$, consistent with the third law of thermodynamics.

The deviation from local equilibrium is quantified by 
the local entropy deficit $\Delta{s}=s_{p}-s_{s}$ shown in the top right panel of Fig.\ \ref{entropy}.
$\Delta{s}$ shows deep blue regions (low entropy deficit) in both the hottest and coldest parts of the system, while the largest entropy deficits
(bright red) occur in the areas at intermediate temperatures.
This may be explained as follows: within elastic transport theory, the local nonequilibrium distribution
function is a linear combination of the distribution functions of the various reservoirs (see Appendix\ \ref{appendLocalEntropy}). 
The entropy deficit is minimal
when this distribution function strongly resembles the equilibrium Fermi-Dirac distribution of one of these reservoirs. 
Conversely, the entropy deficit is maximal when there is a large admixture of both hot and cold electrons without inelastic processes leading to 
equilibration.
Therefore, the hottest and coldest spots show the smallest entropy deficits since
there is very little mixing from the cold reservoir $R1$ and hot reservoir $R2$, respectively, while the regions at intermediate temperatures
have the largest entropy deficitis, and hence are farthest from local equilibrium. 

However, it can be seen that the colder spots are more strongly affected
due to the mixing from the hot reservoir $R2$, while the hotter spots are affected to a lesser extent due to the mixing from the cold reservoir $R1$. 
This reflects the fact that
the distribution function $f_{s}$ deviates much more from the distribution function $f_{1}(\omega)$ with $T_{1}\rightarrow0$ (implying a pure state with 
zero entropy) due to a small admixture of hot electrons from 
$R2$ than is the case for the opposite scenario. In other words, it is easier to increase the entropy deficit
$\Delta{s}$ of a cold spot by adding hot electons (and thus driving it out of equilibrium) than it is the other way round.
The entropy deficit is a good metric to capture such a change in the distribution function, and gives us a per-state `distance' from local equilibrium. 

\section{Conclusions}
\label{sec:conclusions}

We investigated local electronic temperature distributions in nanoscale quantum conductors with one of the reservoirs held at finite temperature
and the other held at or near absolute zero, a problem essentially outside the scope of linear response theory.
The local temperature was defined as that measured by a floating thermoelectric probe, Eq.\ (\ref{equilibrium}).
In particular, we addressed a question motivated
by the third law of thermodynamics: can there be a local temperature
arbitrarily close to absolute zero in a nonequilibrium quantum system?

We obtained local temperatures close
to absolute zero when electrons originating from the finite temperature
reservoir undergo destructive quantum interference.
The local temperature was computed by numerically solving a nonlinear system of
equations [Eqs.\ (\ref{equilibrium}) and (\ref{NumberHeatCurrent})] describing equilibration of a scanning thermoelectric probe with the system, and we
obtain excellent agreement with analytic results [Eqs.\ (\ref{eqn1}), (\ref{eqn2}), and (\ref{polysolution})] derived using the Sommerfeld expansion.
Our conclusion is that a local temperature equal to absolute zero is impossible in a nonequilibrium quantum system, 
but arbitrarily low finite values are possible.

A definition for the local entropy [Eq.\ (\ref{SystemEntropy})]
of a nonequilibrium system of independent fermions was proposed, along with a normalization factor [Eq.\ (\ref{normalization})]
that takes into
account local variations in the density of states.   The local nonequilibrium entropy is always less than or equal to that of a local equilibrium
distribution with the same mean energy and occupancy, and the local entropy deficit was used to quantify the distance from local equilibrium in a 
nanoscale junction with nonlinear thermal bias (Fig.\ \ref{entropy}).
It was shown that the local entropy of the system tends to zero when the probe temperature tends to zero, implying that the local temperature so defined
is consistent with the third law of thermodynamics.

\begin{acknowledgments}
The authors gratefully acknowledge Justin P. Bergfield for assistance in modeling the probe-sample coupling.
This work was
supported by the U.S.\ Department of Energy
(DOE), Office of Science under Award No.\ DE-SC0006699.
\end{acknowledgments}

\appendix
\section{Elastic transport regime}
\label{appendLocalEntropy}

We derive the form of the nonequilibrium distribution function $f_{s}(\omega)$ when the transport is dominated by elastic processes.
We assume a nanostructure connected to $M$ reservoirs, including the probe.
Eq.\ (\ref{Rearraged}) takes the form of Eq.\ (\ref{NumberHeatCurrent}) when the transport is elastic, and we have
\begin{equation}
\begin{aligned}
2\pi\Tr{\Gamma^{p}(\omega)A(\omega)}&\big(f_{s}(\omega)-f_{p}(\omega)\big)\\
&=\sum_{\alpha=1}^{M}\myT_{p\alpha}(\omega)\big(f_{\alpha}(\omega)-f_{p}(\omega)\big).
\end{aligned}
\end{equation}
Now, we wish to rewrite the above equation in terms of the local properties sampled by the probe:
\begin{equation}
\begin{aligned}
&\frac{\Tr{\Gamma^{p}(\omega)A(\omega)}}{\Tr{\Gamma^{p}(\omega)}}\big(f_{s}(\omega)-f_{p}(\omega)\big)\\
&=\sum_{\alpha=1}^{M}\frac{\Tr{\Gamma^{p}(\omega)G^{r}(\omega)\Gamma^{\alpha}(\omega)G^{a}(\omega)}}{2\pi\Tr{\Gamma^{p}(\omega)}}\big(f_{\alpha}(\omega)-f_{p}(\omega)\big),
\label{ElasticDistributionFunction}
\end{aligned}
\end{equation}
where the first factor on the $l.h.s$ is the mean local spectrum $\bar{A}(\omega)$ sampled by the probe, defined by Eq.\ (\ref{meanlocalspec}), and
we used Eq.\ (\ref{TMatrix}) for the elastic transmissions on the $r.h.s$. 
We define the injectivity of a reservoir $\alpha$ sampled by the probe as
\begin{equation}
\rho_{p\alpha}(\omega)=\frac{1}{2\pi}\frac{\Tr{\Gamma^{p}(\omega)G^{r}(\omega)\Gamma^{\alpha}(\omega)G^{a}(\omega)}}{\Tr{\Gamma^{p}(\omega)}},
\label{injectivity}
\end{equation}
for the factors appearing on the $r.h.s$ of Eq.\ (\ref{ElasticDistributionFunction}).
Injectivity of a reservoir $\alpha$ has been previously defined \cite{Gasparian96} as the local partial density of states (LPDOS) associated with
the electrons originating from reservoir $\alpha$ and, due to number conservation, the sum of injectivities of the reservoirs
gives the local density of states (LDOS). We state an equivalent result for the injectivities defined in Eq.\ (\ref{injectivity}) in the
following paragraph.
Before proceeding, we note that the injectivities sampled by the probe, in Eq.\ (\ref{injectivity}), reduces to
the LPDOS for electrons injected by reservoir $\alpha$ when the probe coupling is maximally local,
i.e., $[\Gamma^{p}(\omega)]_{ij}=\Gamma^{p}(\omega)\delta_{in}\delta_{jn}$ and becomes essentially independent of the probe coupling when it is weak.
Eq.\ (\ref{injectivity}) also
extends to $\alpha=p$ and defines the probe injectivity sampled by itself, which becomes negligible in the limit of weak coupling.

It can be shown that the spectrum can be written as \cite{Datta95}
\begin{equation}
A(\omega)=\frac{1}{2\pi}G^{r}(\omega)\Gamma(\omega)G^{a}(\omega),
\label{spec}
\end{equation}
where $\Gamma(\omega)$ is given by
\begin{equation}
\Gamma(\omega)= \sum_{\alpha}\Gamma^{\alpha}(\omega).
\label{Gamma}
\end{equation}
The contribution due to interactions $\Gamma^{\rm{int}}(\omega)$ in Eq.\ (\ref{Gamma}) is missing since the interaction self-energy is Hermitian for elastic processes.
Eqs. (\ref{injectivity}), (\ref{spec}) and (\ref{Gamma})  imply:
\begin{equation}
\sum_{\alpha=1}^{M}\rho_{p\alpha}(\omega)=\bar{A}(\omega).
\label{numberconserve}
\end{equation}
From Eq.\ (\ref{ElasticDistributionFunction}), we write
\begin{equation}
\bar{A}(\omega)\big(f_{s}(\omega)-f_{p}(\omega)\big)=\sum_{\alpha=1}^{M}\rho_{p\alpha}(\omega)\big(f_{\alpha}(\omega)-f_{p}(\omega)\big)
\end{equation}
and Eq.\ (\ref{numberconserve}) implies
\begin{equation}
\bar{A}(\omega)f_{s}(\omega)=\sum_{\alpha=1}^{M}\rho_{p\alpha}(\omega)f_{\alpha}(\omega).
\label{ElasticPenultimate}
\end{equation}
Finally,
$f_{s}(\omega)$ can be written as
\begin{align}
f_{s}(\omega)&=\sum_{\alpha=1}^{M}\frac{\rho_{p\alpha}(\omega)}{\bar{A}(\omega)}f_{\alpha}(\omega).\\
\therefore \ \ 0\leq f_{s}(\omega)&\leq\sum_{\alpha=1}^{M}\frac{\rho_{p\alpha}(\omega)}{\bar{A}(\omega)}\\
0\leq f_{s}(\omega)&\leq1,
\label{ElasticFinal}
\end{align}
where we used Eq.\ (\ref{numberconserve}) and the fact that the Fermi-Dirac distributions satisfy $0\leq f_{\alpha}(\omega)\leq1$.
The nonequilibrium distribution function $f_{s}(\omega)$ is thus a linear combination of the Fermi-Dirac distributions of the reservoirs, and
Eq.\ (\ref{ElasticFinal})
leads to an unambiguous definition of the local entropy for a nonequilibrium system given by Eq.\ (\ref{SystemEntropy}).

\section{Model of probe-sample coupling}
\label{TipSampleCoupling}

The scanning thermoelectric probe is modeled as an atomically sharp Au tip operating in the tunneling regime. The probe tunneling-width matrices may be described in general
as $\Gamma^{p}_{nm}(\omega)=2\pi V_{n}V_{m}^{*}\rho_{p}(\omega)$, where $\rho_{p}(\omega)$ is the local density of states of the apex atom
in the probe electrode and $V_{m}$,$V_{n}$ are the tunneling matrix elements between the quasi-atomic wavefunctions of the apex atom in
the electrode and the $m^{th}$, $n^{th}$ $\pi$-orbitals in the molecule. We consider the Au tip to be dominated by the s-orbital character and neglect
all other contributions. The probe-system coupling is also treated within the broad-band approximation. The tunneling-width matrix $\Gamma^{p}$ describing
the probe-system coupling is
in general non-diagonal, and is calculated using the methods highlighted in Ref.\ \onlinecite{Chen93}.

\section{Exact solution}
\label{append:ExactSolution}

The exact solution to the Eq. (\ref{QNodeHeatCurrent}) with $T_{1}\rightarrow0$ is
\begin{equation}
T_{p}=T_{2}\bigg(\frac{\sqrt{1 + 4\lambda_{1}^{2}}-1}{2\lambda_{1}}\bigg)^{\frac{1}{2}},\nonumber
\end{equation}
where $\lambda_{1}$ is defined in Eq.\ (\ref{lambda}) and simplifies to
\begin{equation}
\lambda_{1}=\frac{7\pi^{2}}{20}\frac{\myT_{p2}^{(2)}(k_{B}T_{2})^{2}}{\myT_{p1}},\nonumber
\end{equation}
a factor that appears in Eq.\ (\ref{eqn2}).

\bibliography{refs,refs_graphene}

\end{document}